\documentclass[lettersize,journal]{IEEEtran}
\usepackage{amsmath,amsfonts}
\usepackage{algorithmic}
\usepackage{algorithm}
\usepackage{array}
\usepackage[caption=false,font=normalsize,labelfont=sf,textfont=sf]{subfig}
\usepackage{textcomp}
\usepackage{stfloats}
\usepackage{url}
\usepackage{verbatim}
\usepackage{graphicx}
\usepackage{cite}
\newtheorem{lemma}{Lemma}
\newtheorem{theorem}{Theorem}
\newcommand{\at}[2][]{#1|_{#2}}

\begin{document}

\title{Optimizing Age of Information in Random Access Networks: A Second-Order Approach for Active/Passive Users}

\author{Siqi Fan,~\IEEEmembership{Graduate Student Member,~IEEE,} Yuxin Zhong,~\IEEEmembership{Graduate Student Member,~IEEE,} \\ I-Hong Hou,~\IEEEmembership{Senior Member,~IEEE,} Clement K Kam,~\IEEEmembership{Senior Member,~IEEE} 
        % <-this % stops a space
\thanks{Siqi Fan and I-Hong Hou are with the Department of Electrical and Computer Engineering, Texas A\&M University, 188 Bizzell St, College Station, TX 77843 (e-mail: siqifan@tamu.edu; $\;$ ihou@tamu.edu).

Yuxin Zhong is with the School of Electrical and Computer Engineering, Georgia Institute of Technology, 777 Atlantic Drive NW, Atlanta, Georgia (e-mail: yzhong332@gatech.edu). Her work was done in her study in Texas A\&M university.

Clement K Kam is with the U.S. Naval Research Laboratory, Washington, USA (e-mail: clement.kam@nrl.navy.mil).}% <-this % stops a space
\thanks{Manuscript received August 30, 2023; revised April 14, 2024.}}



\maketitle

\begin{abstract}
In this paper, we study the moments of the Age of Information (AoI) for both active and passive users in a random access network. In this network, active users broadcast sensing data, while passive users detect in-band radio activities from out-of-network devices, such as jammers. Collisions occur when multiple active users transmit simultaneously. Passive users can detect radio activities only when no active user transmits. Each active user's transmission behavior follows a Markov process. We aim to minimize the weighted sum of any moments of AoI for both user types. To achieve this, we employ a second-order analysis of system behavior. Specifically, we characterize an active user's transmission Markov process using its mean and temporal variance. We show that any moment of the AoI can be approximated by a function of these two parameters. This insight enables us to analyze and optimize the transmission Markov process for active users. We apply this strategy to two different random access models. Simulation results show that policies derived from this strategy outperform other baseline policies. 
\end{abstract}


\section{Introduction}
\IEEEPARstart{R}{ecently}, there has been a significant increase in the use of time-sensitive applications such as Internet of Things, vehicular networks, sensor networks, and drone systems for surveillance and reconnaissance. While traditional research has focused on optimizing the reliability and throughput of communication, these efforts often fall short in meeting the demands for fresh data. To address this, a performance metric known as age-of-information (AoI) was introduced \cite{kaul2011minimizing} and has received increasing attention in recent literature. 

In this paper, we focus on AoI performance in a random access network where active users, such as video sensors, broadcast their sensed data. When interfering active users transmit at the same time, a collision occurs, and all transmissions fail.

Additionally, we consider that the network has silent observers, referred to as passive users, who share the same channel as active users. Passive users aim to observe out-of-network radio activities in the same channel but do not make any transmissions. Examples of passive users include sensors detecting malicious jammers, receivers of satellite communications, and radio telescopes tracking celestial objects.

As AoI depends on previous successful transmissions, memoryless random access algorithms, such as slotted ALOHA, are not suitable for optimizing AoI. Therefore, we consider that each active user follows a Markov process to determine its transmission activities. Furthermore, instead of only considering the mean of AoI, we consider the moments of AoI, which can offer insights into the variance and the maximum value of AoI.

The goal of this paper is to minimize the weighted sum of any moment of AoI for both active users and passive users. There are several challenges that need to be addressed. First, since the transmission model of an active user is Markovian instead of i.i.d., the temporal correlation of its transmission activities needs to be explicitly taken into account. Second, active users and passive users have different behavior, which leads to different definitions of AoI between the two types of users. Third, most existing studies on moments of AoI, such as \cite{inoue2017stationary,moltafet2022aoi, moltafet2022moment}, focus only on simple queuing systems and are not applicable to random access networks. Finally, optimizing only the mean of AoI in random access networks may not be sufficient in certain cases. For instance, in scenarios where we require the AoI to be stable for accurate and precise analysis and prediction, optimizing the variance of AoI becomes essential. This can be achieved by calculating the second moment of AoI. However, most existing studies on AoI in random access networks only focus on the first moment of AoI and can only obtain the optimal solution in the asymptotic sense.

\begingroup\renewcommand\thefootnote{\textsection}
\footnotetext{The original concept of applying second-order analysis to AoI in random access networks was introduced in our preliminary conference paper \cite{fan2023minimizing}. That study focused solely on the performance of a two-state model, which demonstrated performance similar to that of optimal ALOHA in simulations.  In the present paper, we extend this work by introducing a new model that has been shown to significantly reduce AoI in simulations. We also provide closed-form approximation expressions for evaluating its performance and discuss how to obtain optimal parameter settings.}

To address above challenges and solve the optimization problem, we propose to investigate the system behavior through a second-order analysis. This paper's key contributions are as follows\textsuperscript{\textsection}:
\begin{itemize}
    \item We demonstrate that the moments of AoI for both active and passive users can be approximated by functions of the mean and temporal variance of the delivery process. Additionally, we provide closed-form expressions for these approximations.
    \item In the context of a two-state model for active users, we present expressions detailing the mean variances of the delivery process based on state transmission probabilities. By identifying specific properties within the expressions of moments of AoI for both user types, we find that the optimal strategy of the approximated objective function for an active user, under some minor conditions, is to become silent immediately after one attempt of transmission. The simulation results show that this proposed optimal strategy for the approximated objective function closely aligns with the optimal strategy in practice.
    \item Leveraging the insights from the two-state model, we propose a novel model that can greatly increase the AoI performance. Then, we detail the moments of AoI using functions of two model parameters and demonstrate that finding optimal parameter settings for approximated AoI is achievable through two simple line searches.
    \item Through various simulations, we validate the accuracy of our AoI estimations, efficiency of using proposed analysis strategy instead of real simulations, and compare our model against other state-of-the-art algorithms, including slotted ALOHA, optimal ALOHA, Age-Threshold ALOHA, and a newly proposed pre-assigned sequence algorithm. The outcomes indicate that our approximations are accurate, and our proposed model outperforms competing algorithms in performance. This superior performance is even more remarkable considering that the pre-assigned sequence algorithm employs a centralized server to dictate transmission sequences for active users, in contrast to our completely distributed algorithm.
\end{itemize}

The rest of the paper is organized as follows. In Section~\ref{sec:related}, we delve into the existing literature regarding AoI. In Section~\ref{sec:system}, we present our system model and define the optimization problem. Section~\ref{sec:model} develops the expressions for moments of AoI for both active and passive users based on a second-order model. Section~\ref{sec:two_states} presents an analysis of a basic two-state model and solves the optimal settings for this model. Building on the optimal solutions of the two-state model, we introduce and analyze a novel model in Section~\ref{sec:wag}. We conduct various simulations to compare our model with other algorithms in Section~\ref{sec:sim}. Next, Section~\ref{sec:future} describes the future work. Lastly, Section~\ref{sec:conclusion} concludes the paper. 

%%%%%%%%%%%%%%%%%%%%%%%%%%%%%%%%%%%%%%%%%%%%%%%%%%%%%%%%%%%%%%%%%%%%%%%%%%%%%%%

\section{Related Work}
\label{sec:related}

The concept of AoI has attracted significant attention in the research community. Kaul \textit{et al.} were the first to introduce AoI in their work \cite{kaul2011minimizing}, where they seek to quantify the application-to-application delay observed in a vehicle's state. They further expand on this in \cite{kaul2012real}, presenting general methods for calculating AoI. Consequently, there is a surge in research on AoI across various application domains.

Initially, much of the research on AoI optimization concentrated on queuing problems. In these studies, \cite{bedewy2017age, najm2019content, yates2018status, yang2021understanding} focus on the first-come-first-served and last-come-first-served policies. Studies \cite{chen2016age, kosta2017age, moltafet2020average, huang2015optimizing} analyze AoI in the M/M/1 or M/G/1 queuing models. From another perspective, Wang \textit{et al.} \cite{wang2022useful} study a system with queues in both forward and backward directions, while Zou \textit{et al.} \cite{zou2021optimizing} consider both transmit and computation queues. To further enrich the field, \cite{inoue2017stationary, moltafet2022aoi, moltafet2022moment} formulate the moments of AoI across various queuing models, providing insights into the distribution dynamics of AoI. However, all of these studies focus on single source situations, which are not applicable in multi-source random access networks.

Subsequently, discussions have emphasized AoI optimization via centralized transmission scheduling policies for multi-source systems. These discussions generally bifurcate into two categories: transmitting updated information from a central server to users, and users sending their updated information to the central server. In the realm of the former, Hsu \textit{et al.} \cite{hsu2017age} and Banerjee \textit{et al.} \cite{banerjee2020fundamental} minimize AoI in broadcast networks and cellular wireless networks, respectively. Turning to the latter category, Sun \textit{et al.} \cite{sun2018age} and Kadota \textit{et al.} \cite{kadota2018scheduling} study different scheduling policies. Using Whittle’s index techniques, a series of studies \cite{sun2019closed, maatouk2020asymptotically, kriouile2021minimizing} provide various insights and closed-form expressions for AoI optimization in different contexts. With specific channel considerations in mind, Guo \textit{et al.} \cite{guo2022theory} and Yao \textit{et al.} \cite{yao2022age} apply different mathematical techniques to optimize AoI for Gilbert-Elliott channels. Additionally, several studies also highlight trade-offs, such as those by Jaiswal \textit{et al.} \cite{jaiswal2021minimization} and Chen \textit{et al.} \cite{chen2021timely}. Using the Stochastic Hybrid Systems tool, Yates \textit{et al.} \cite{yates2018age} and Maatouk \textit{et al.} \cite{maatouk2022analysis} provide a comprehensive analysis of status update systems. Besides, Bedwy \textit{et al.} \cite{bedewy2021optimal} introduce a mathematical strategy to optimize the Age of Information in systems by managing generation, transmission, and queuing via a centralized controller. Noting that centralized policies can introduce substantial overhead and delays, these studies are unsuitable for low-powered and delay-sensitive applications like the Internet of Things. Although recent studies on pre-assigned sequence policies, such as that by Liu \cite{liu2023age}, bypass the need for synchronization and feedback, they still rely on a central server for sequence assignments. Nevertheless, these studies require the system to have only one scheduler that controls the scheduling policies of all transmissions, rendering them impractical for broadcasting scenarios.

In recent years, the study of AoI optimization in a distributed setup has gained increased attention. For instance, Yates \textit{et al.} \cite{yates2020age} derive the AoI for an unslotted, uncoordinated, unreliable multiple access collision channel under certain specific settings. Taking into account the access to transmission results or the status of other users, various enhanced algorithms have been proposed. For example, Chen \cite{chen2020age} and Yavascan \cite{yavascan2021analysis} propose optimal solutions for different age-based slotted ALOHA-type algorithms. Yates \cite{yates2017status} presents a slotted ALOHA-like random access policy. Pan \cite{pan2022age} introduces a collision resolution procedure for users whose transmissions encounter collisions. Employing game theory, He \cite{he2022decentralized} and Saurav \cite{saurav2021game} propose different scheduling mechanisms. However, none of above works consider the coexistance of passive user, or analyze the moment of AoI.

%%%%%%%%%%%%%%%%%%%%%%%%%%%%%%%%%%%%%%%%%%%%%%%%%%%%%%%%%%%%%%%%%%%%%%%%%%%%%%%%%

\section{System Settings}
\label{sec:system}

We consider a random access broadcast wireless system with one channel and two types of users, active users and passive users. Each active user monitors a particular field of interest and uses the wireless channel to broadcast its surveillance information to its neighboring receivers. Passive users do not make any wireless transmissions and instead monitor the radio activities in the wireless channel. For example, in a battlefield scenario, an active user can be a surveillance drone that monitors a certain area on the battlefield and broadcasts its video feed to all nearby units. A passive user can be a signal detector that aims to identify and locate enemy jammers or communication devices operating in the same channel. Fig.~\ref{fig:example} presents an example of the system, featuring two active user groups and three passive user detectors. Each group comprises four active user drones along with their corresponding receivers.

\begin{figure}[!th]
\centering
\includegraphics[scale=0.23]{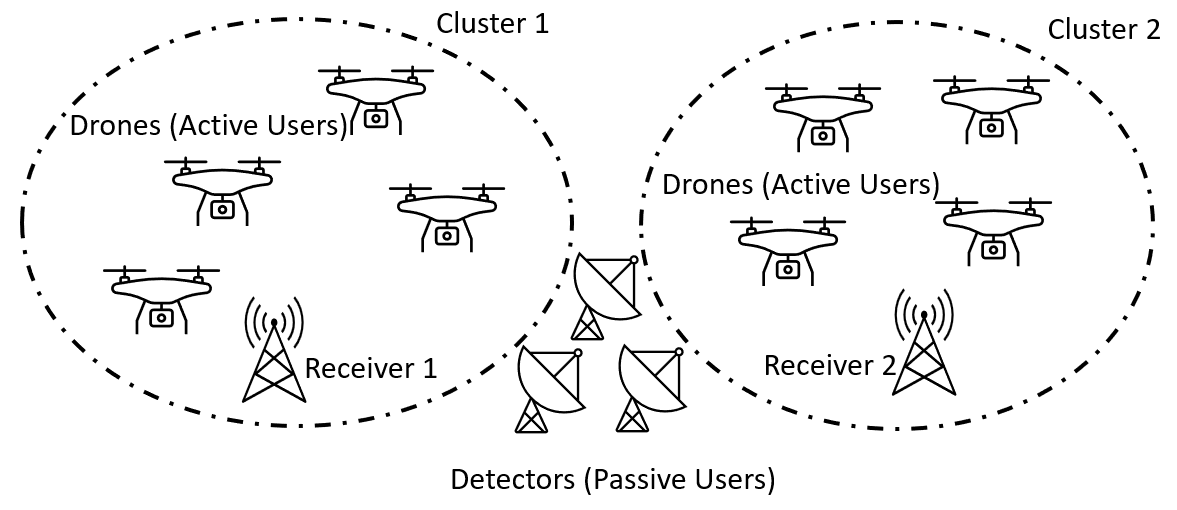}
\caption{An example of active users (drones) and passive users (detectors).}
\label{fig:example}
\end{figure}

Next, we discuss the interference relationship among users. We assume that active users are grouped into $C$ clusters and each cluster has $N$ ($N\geq 2$) active users. Active users in the same cluster interfere with each other but do not suffer from interference from other clusters. An active user can successfully deliver a packet if it is the only device in the cluster that is transmitting. If multiple active users in the same cluster transmit at the same time, then a collision occurs and none of the packets are received. Since all transmissions are broadcasts, which do not incur ACKs in most wireless standards, an active user does not know whether its transmissions suffer from collisions. On the other hand, since passive users need to detect enemy devices that are potentially far away, we assume that a passive user is influenced by all active users and it can only detect radio activities if no active user is transmitting.

We assume that each active user follows an ergodic Markov process with parameter vector $\varsigma$ to determine its transmission activities. For example, we can consider a simple two-state Markov process as shown in Fig.~\ref{fig:two_states}, where the active user only transmits when it is in state TX. In this example, $\varsigma$ is defined as the vector $[r,s]$.

\begin{figure}[!th]
\centering
\includegraphics[scale=0.23]{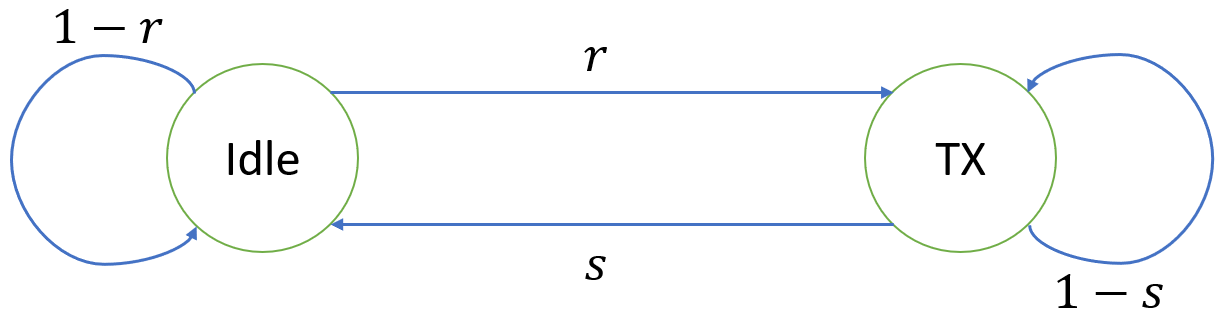}
\caption{The two-state model.}
\label{fig:two_states}
\end{figure}

The performance of each user is determined by its associated AoI. We assume that time is slotted and the duration of one time slot is the time needed to transmit one packet. Each active user generates a new surveillance packet in each slot. Hence, the AoI of an active user $n$ at time $t$ is defined by the following recursion:
\begin{align*}
    AoI_{n,t} := \begin{cases}
                    1, & \text{if users $n$ successfully delivers} \\
                    & \text{a packet at time $t$,} \\
                    AoI_{n,t-1} + 1, & \text{otherwise}.
                \end{cases}
\end{align*}
Since passive users can only detect radio activities when no active users are transmitting, the AoI of passive user $\varrho$ at time $t$ is defined by
\begin{align*}
    AoI_{\varrho,t} := \begin{cases}
                    1, \; & \text{if no active user transmits,} \\
                    AoI_{\varrho,t-1} + 1, \; & \text{otherwise}.
                \end{cases}
\end{align*}

We aim to minimize the moments of AoI for both active users and passive users. Considering that all active users are identical, as well as all passive users, we focus on one active user and one passive user to construct the objective function. Specifically, let $AoI_a$ be the random variable representing an active user's AoI in a steady state, $AoI_p$ be the random variable representing a passive user's AoI in a steady state, both under parameter vector $\varsigma$. For a fixed integer $z$ and a fixed real number $w\in[0,1]$, we aim to find the optimal $\varsigma$ that solve the following optimization problem:
\begin{align}
    \min \; & F(\varsigma) := w \sqrt[z]{E[AoI_a^z]} + (1-w) \sqrt[z]{E[AoI_p^z]}, \label{opt_problem}
\end{align}
where $E[\cdot]$ is the expectation function.

It should be noted that the system is reduced to only considering active users when $w=1$.

%%%%%%%%%%%%%%%%%%%%%%%%%%%%%%%%%%%%%%%%%%%%%%%%%%%%%%%%%%%%%%%%%%%%%%%%%%%%%%%

\section{The Second Order Model for AoI}
\label{sec:model}

A key challenge in solving the optimization problem (\ref{opt_problem}) is that it is hard to express $E[AoI_a^z]$ and $E[AoI_p^z]$ as functions of Markov process parameters $\varsigma$. We propose to adopt the framework of second-order network optimization in Guo \textit{et al.} \cite{guo2022theory} to address this challenge. The framework in \cite{guo2022theory} is only applicable to centrally scheduled systems and the mean of AoI. In this section, we will extend this framework to model random access networks and moments of AoI. The basic idea is to characterize the transmission process based on the mean and temporal variance of successful communications. By utilizing these means and temporal variances, we can then approximate the AoI.

We first formally define the second-order model that describes the performance of both active and passive users. For an active user $n$, let $S_n(t)$ be the indicator function that $n$ successfully delivers a packet at time $t$, that is, $n$ is the only active user in its cluster that transmits at time $t$. We call $[S_n(1), S_n(2), \dots]$ the \emph{delivery process} of $n$. Due to symmetry, the delivery processes of all active users follow the same distribution. We then define the mean and temporal variance of the delivery process of an active user by
\begin{align*}
    m_a &:= \lim_{T\xrightarrow{} \infty } \frac{\sum_{t=1}^{T}S_n(t)}{T}, \\
    v_a^2 &:= \lim_{T\xrightarrow{} \infty} E[( \frac{\sum_{t=1}^{T}S_n(t)-T m_a}{\sqrt{T}})^2],
\end{align*}
respectively.

For a passive user $\varrho$, we let $S_{\varrho}(t)$ be the indicator function that no active user transmits, and hence the passive user can monitor radio activities without interference, at time $t$. We call $[S_{\varrho}(1), S_{\varrho}(2), \dots]$ the \emph{passive detection process}. The mean and temporal variance of the passive detection process are defined as
\begin{align*}
    m_p &:= \lim_{T\xrightarrow{} \infty } \frac{\sum_{t=1}^{T}S_{\varrho}(t)}{T}, \\
    v_p^2 &:= \lim_{T\xrightarrow{} \infty} E[( \frac{\sum_{t=1}^{T}S_{\varrho}(t)-T m_p}{\sqrt{T}})^2],
\end{align*}
respectively.

Next, we demonstrate that the moments of AoI for both active and passive users can be approximated as functions of $(m_a, v_a^2)$ and $(m_p, v_p^2)$, respectively. Since the analysis pertaining to active users is analogous to that of passive users, we will primarily concentrate on detailing the derivation of the moments of AoI for the active users. Following a similar approach, the expressions for the moments of AoI for passive users can be subsequently derived.

Let $t_i$ be the $i$-th time that an active user $n$ successfully delivers its packet, and define $l_n(i):=t_i-t_{i-1}$. For active user $n$, its AoI from $t_{i-1}$ to $t_i-1$ is from $1$ to $l_n(i)$. Therefore, the summation of $z$-th moment of AoI from $t_{i-1}$ to $t_i-1$ is $\sum_{k=1}^{l_n(i)} k^z$. Let $L_n$ represent the total number of deliveries for active user $n$. By summing over all $l_n(i)$, we obtain the sum of all $z$-th moment of AoI for active user $n$, which is given by $\sum_{i=1}^{L_n} \sum_{k=1}^{l_n(i)} k^z$. Then, by dividing this sum by the total time $\sum_{i=1}^{L_n} l_n(i)$ and taking the expectation, we have:
\begin{align*}
    E[AoI_a^z] = E\Big[\lim_{L_n\xrightarrow{} \infty}\frac{\sum_{i=1}^{L_n}\sum_{k=1}^{l_n(i)}k^z}{\sum_{i=1}^{L_n}l_n(i)}\Big]
\end{align*}
By using Faulhaber’s formula, we have 
\begin{align}
    & E[AoI_a^z]  =  E\Big[\lim_{L_n\xrightarrow{} \infty}\frac{\sum_{i=1}^{L_n}1}{\sum_{i=1}^{L_n}l_n(i)}\Big(\frac{1}{\sum_{i=1}^{L_n}1}\sum_{i=1}^{L_n} \big( \frac{l_n(i)^{z+1}}{z+1} \nonumber  \\
    & \qquad + \frac{l_n(i)^z}{2} + \sum_{\kappa=2}^{z} \frac{B_{\kappa}}{\kappa!}\frac{z!l_n(i)^{z-\kappa+1}}{(z-\kappa+1)!} \big) \Big)\Big] \nonumber \\
    & = \frac{1}{E[l_n(i)]}\big(\frac{E[l_n(i)^{z+1}]}{z+1} + \frac{E[l_n(i)^z]}{2} \nonumber \\
    & \qquad + \sum_{\kappa=2}^z\frac{B_{\kappa}}{\kappa!}\frac{z!E[l_n(i)^{z-\kappa+1}]}{(z-\kappa+1)!} \big),
\end{align}
where $B_{\kappa}$ is the Bernoulli number. 

Next, we derive the expressions for $E[l_n(i)^{\kappa}]$ for any $\kappa$. It is important to note that $l_n(i)$ is defined as the time interval between the $(i-1)$-th and $i$-th occurrences when $S_n(t)=1$. Since $S_n(t)$ is hard to track, we construct an alternative sequence $S'_n(t)$ that has the same mean $m_a$ and temporal variance $v_a^2$ as $S_n(t)$ to assist us in approximating $l_n(i)$. This sequence is formally defined as follows. Let $D_n(t)$ be a Brownian motion random process with mean $m_a$ and variance $v_a^2$, then the sequence $\{S'_n(1), S'_n(2), \dots\}$ is defined as:
\begin{align}
    S'_n(t) =  \begin{cases}
                    1, \; & \text{if $D_n(t)-D_n(t^-) \geq 1$,} \\
                    0, \; & \text{otherwise,}
                \end{cases}
    \label{eq:def_s'}
\end{align}
where $t^-=\max\{\tau|\tau<t, S'_n(\tau)=1\}$. Intuitively, in the alternative sequence, we assume that a packet delivery occurs every time $D_n(t)$ increases by one since the last packet delivery. Hence, $\sum_{\tau=1}^t S_n'(\tau)$ is close to $D_n(t)$, and the process $[S'_n(1), S'_n(2), ...]$ has the mean $m_a$ and temporal variance $v_a^2$.

Let $l_n'(i):=t'_{n,i} - t'_{n,i-1}$, where $t'_{n,i}$ is the $i$-th time that $S'_n(t)=1$. Since $S'_n(t)$ has the same mean and temporal variance as $S_n(t)$, we propose to approximate $E[l_n(i)^{\kappa}]$ by $E[l'_n(i)^{\kappa}]$. Then, $E[AoI_a^z]$ can be approximated using $E[l'_n(i)^{\kappa}]$. Denote the approximated active user’s AoI as $E[\widetilde{AoI_a}]$, we have
\begin{align*}
    E[\widetilde{AoI_a}^z] = & \frac{1}{E[l'_n(i)]}(\frac{E[l'_n(i)^{z+1}]}{z+1} + \frac{E[l'_n(i)^z]}{2} \nonumber\\
        & + \sum_{{\kappa}=2}^{z}\frac{B_{\kappa}}{{\kappa}!}\frac{z! E[l'_n(i)^{z-{\kappa}+1}]}{(z-{\kappa}+1)!}).
\end{align*}

Moreover, $l'_n(i)$ can be regarded as the time needed for $D_n(t)$ to increase by $1$, which is equivalent to the first-hitting time for the Brownian motion random process with level $1$. Thus, $l_n'(i)$ follows the inverse Gaussian distribution $IG(\frac{1}{m_a},\frac{1}{v_a^2})$ (\cite{schrodinger1915theorie,folks1978inverse}). By using the moment generating function of inverse Gaussian distribution in \cite{krishnamoorthy2006handbook}, we have
\begin{align*}
    E[l'_n(i)^{\kappa}] = \frac{1}{m_a^{\kappa}}\sum_{\zeta=0}^{{\kappa}-1}\frac{({\kappa}-1+\zeta)!}{\zeta!({\kappa}-1-\zeta)!}(\frac{2m_a}{v_a^2})^{-\zeta}.
\end{align*}

For a passive user $\varrho$, following the same strategy, let $\Bar{t}_i$ be the $i$-th time that $S_{\varrho}(t)=1$. Define $l_j(i):=\Bar{t}_i-\Bar{t}_{i-1}$, and let $l'_j(i)$ follows the inverse Gaussian distribution $IG(\frac{1}{m_p},\frac{1}{v_p^2})$, approximating $l_j(i)$. Denote the approximated passive user’s AoI as $E[\widetilde{AoI_p}^z]$. Combining the above derivations, we can approximate the solution for optimization problem (\ref{opt_problem}) by solving the following optimization problem:
\begin{align}
    \min \tilde{F}(\varsigma) := w \sqrt[z]{E[\widetilde{AoI_a}^z]} + (1-w) \sqrt[z]{E[\widetilde{AoI_p}^z]}.
    \label{eq:relax_opt_problem}
\end{align}
The expressions for $E[\widetilde{AoI_a}^z]$ and $E[\widetilde{AoI_p}^z]$ are given by the following theorem:
\begin{theorem}
    Let $B_k$ be the Bernoulli number, then
    \begin{align}
        E[\widetilde{AoI_a}^z] = & \frac{1}{E[l'_n(i)]}(\frac{E[l'_n(i)^{z+1}]}{z+1} + \frac{E[l'_n(i)^z]}{2} \nonumber\\
        & + \sum_{{\kappa}=2}^{z}\frac{B_{\kappa}}{{\kappa}!}\frac{z! E[l'_n(i)^{z-{\kappa}+1}]}{(z-{\kappa}+1)!}),
        \label{eq:active_aoi}
    \end{align}
    where
    \begin{align}
         &E[l'_n(i)^{{\kappa}}] = \frac{1}{m_a^{\kappa}}\sum_{\zeta=0}^{{\kappa}-1}\frac{({\kappa}-1+\zeta)!}{\zeta!({\kappa}-1-\zeta)!}(\frac{2m_a}{v_a^2})^{-\zeta},
         \label{eq:active_moment}
    \end{align}
    and
    \begin{align}
        E[\widetilde{AoI_p}^z] =& \frac{1}{E[l'_j(i)]}(\frac{E[l'_j(i)^{z+1}]}{z+1} + \frac{E[l'_j(i)^z]}{2} \nonumber\\
        & + \sum_{{\kappa}=2}^{z}\frac{B_{\kappa}}{{\kappa}!}\frac{z! E[l'_j(i)^{z-{\kappa}+1}]}{(z-{\kappa}+1)!}),
        \label{eq:passive_aoi}
    \end{align}
    where
    \begin{align}
        E[l'_j(i)^{{\kappa}}] = \frac{1}{m_p^{\kappa}}\sum_{\zeta=0}^{{\kappa}-1}\frac{({\kappa}-1+\zeta)!}{\zeta!({\kappa}-1-\zeta)!}(\frac{2m_p}{v_p^2})^{-\zeta}.
        \label{eq:passive_moment}
    \end{align}
    \label{theorem:aoi_z}
    $\hfill\Box$
\end{theorem}
For example, we have 
\begin{align*}
    E[AoI_a] \approx E[\widetilde{AoI_a}] = \frac{1}{2}(\frac{v_a^2}{m_a^2}+\frac{1}{m_a})+\frac{1}{2},
\end{align*}
and
\begin{align*}
    E[AoI_a^2] \approx E[\widetilde{AoI_a}^2] = \frac{v_a^4}{m_a^4}+\frac{v_a^2}{m_a^3}+\frac{3v_a^2+2}{6m_a^2}+\frac{1}{2m_a}+\frac{1}{6}.
\end{align*}

According to Theorem~\ref{theorem:aoi_z}, even if we do not know the intricate details of the Markov process that governs the transmissions of active users, we can still approximate AoI for both active and passive users. Such an approximation is feasible provided that we know $m_a$, $v_a^2$, $m_p$, and $v_p^2$.

The remaining challenge in optimizing the moments of AoI lies in expressing the means and temporal variances as functions of the Markov process parameters, and then resolving the optimization problem (\ref{eq:relax_opt_problem}) to obtain the approximated optimal solution for (\ref{opt_problem}). This is closely tied to the Markov process which governs active user transmissions. In the next two sections, we will study two different kinds of Markov processes, namely, the two-state model and the Wait-and-Go (WaG) model. For both models, we will derive the closed-form expressions of $m_a$, $v_a^2$, $m_p$, and $v_p^2$, in terms of the parameter vector $\varsigma$. Subsequently, we'll employ these expressions to determine the optimal $\varsigma$ through simple grid searches. In the following analysis, the optimal solutions we discuss pertain to proposed approximation of the weighted AoI. Although the theoretical tightness is not guaranteed, we show through simulations that this approximation and its corresponding optimal solutions perform well in practice.

%%%%%%%%%%%%%%%%%%%%%%%%%%%%%%%%%%%%%%%%%%%%%%%%%%%%%%%%%%%%%%%%%%%%%%%%%

\section{Two-state Markov Model}
\label{sec:two_states}

To explore the transmission model of active users, we begin our analysis with the basic two-state Markov model shown in Fig.~\ref{fig:two_states}. In this model, each active user can exist in one of two states: the TX state or the Idle state. The active user will transmit a packet at time $t$ if and only if it is in the TX state at that time. If an active user is in the TX state at time $t$, it will transition to the Idle state at time $t+1$ with a probability $s$. Conversely, if an active user is in the Idle state at time $t$, it will transition to the TX state at time $t+1$ with a probability $r$. In this model, $\varsigma$ is the vector $[r,s]$.

It's noteworthy that this model is a generalization of the ALOHA network. Specifically, if $r+s=1$, then an active user will transmit in each time slot with a probability $r$ independently.

\subsection{Mean and Temporal Variance}
\label{subsec:tx_mv}

Based on the second-order model discussed in Section~\ref{sec:model}, the critical parameters characterizing the transmissions are the means and temporal variances. Thus, it is essential to determine the means and temporal variances of both active and passive users in this basic two-state Markov model.

When the system reaches a steady state, the mean for any active user, denoted as $m_a$, is the probability that only this active user in its cluster is in the TX state. The mean for passive users, denoted as $m_p$, corresponds to the probability that no active user in any cluster is in the TX state. Let $\lambda:=\frac{r}{r+s}$ and $\theta := 1-r-s$. The steady-state probability of an active user being in TX state is $\frac{r}{r+s}=\lambda$, while the probability of being in Idle state is $1-\lambda$. Hence, the means can be calculated by:
\begin{align*}
    m_a = \lambda(1-\lambda)^{N-1}, \qquad m_p = (1-\lambda)^{CN}.
\end{align*}

Next, we derive the temporal variances. Since the derivation techniques for $v_a^2$ and $v_p^2$ are similar, we only demonstrate the derivation of $v_a^2$. Our derivation is based on the central limit theorem of the Markov process:
\begin{align}
    v_a^2 = Var(S_n(1)) + 2\sum_{k=1}^{\infty} Cov(S_n(1),S_n(k+1)).
    \label{eq:va1}
\end{align}
Since $S_n(1)$ is a Bernoulli random variable with mean $m_a$, we have
\begin{align}
    Var(S_n(1))&=E[S_n(1)]-E[S_n(1)]^2 = m_a -m_a^2.
    \label{eq:va2}
\end{align}
Moreover,
\begin{align}
    &Cov(S_n(1), S_n(k+1)) \nonumber\\
    &= E[S_n(1)S_n(k+1)]-E[S_n(1)]E[S_n(k+1)] \nonumber\\
    &=(Prob(S_n(k+1)=1|S_n(1)=1)-E[S_n(k+1)]) E[S_n(1)] \nonumber\\
    & = (Prob(S_n(k+1)=1|S_n(1)=1) - m_a)m_a,
    \label{eq:va3}
\end{align}
To obtain $Prob(S_n(k+1)=1|S_n(1)=1)$, we note that $S_n(t)$ is $1$ if and only if active user $n$ transmits at time $t$ and all the other $N-1$ interfering active users are idle at time $t$. Since all active users follow the same transmission strategy, we can define $G(k) = Prob(x_{n,k}=1|x_{n,1}=1)$ and $\Bar{G}(k) = Prob(x_{n,k}=0|x_{n,1}=0)$. Then,
\begin{align}
    Prob (S_n(k+1)=1|S_n(1)=1) = G(k+1) \Bar{G}(k+1)^{N-1}.
    \label{eq:va4}
\end{align}

It remains to find $G(k)$ and $\Bar{G}(k)$. Consider the relationship between $G(k)$ and $G(k-1)$, we derive:
\begin{align*}
    & G(k) = Prob(x_{n,k}=1|x_{n,1}=1) \\
    &= G(k-1) Prob(x_{n,k}=1|x_{n,k-1}=1) \\
    &\quad + (1-G(k-1))Prob(x_{n,k}=1|x_{n,k-1}=0) \\
    &= G(k-1)(1-s) + (1-G(k-1))r = r + \theta G(k-1),
\end{align*}
for $k>1$, and $G(1)=1$. Then, making the summation $\sum_{i=2}^{k}\theta^{k-i} G(i)$, we get:
\begin{align}
    G(k) &= \sum_{i=2}^{k}\theta^{k-i}r + \theta^{k-1}G(1) = \frac{r}{r+s}+\frac{s}{r+s}\theta^{k-1} \nonumber\\
    & = \lambda + (1-\lambda)\theta^{k-1}.
    \label{eq:va5}
\end{align}
Similarly, we can derive:
\begin{align}
    \Bar{G}(k) = 1-\lambda+\lambda \theta^{k-1}.
    \label{eq:va6}
\end{align}

Putting Equations (\ref{eq:va1})--(\ref{eq:va6}) together, we have:
\begin{align*}
    & v_a^2 =  2\sum_{k=1}^{\infty}((\lambda + (1-\lambda) \theta^{k})(1-\lambda+\lambda \theta^{k})^{N-1} -m_a) m_a \\
    &\quad + m_a - m_a^2.
\end{align*}

In summary, we have established the following theorem:
\begin{theorem}
    For the two-state model, the means and temporal variances are:
    \begin{align*}
        m_a &= \lambda (1-\lambda)^{N-1}, \quad m_p = (1-\lambda)^{CN}, \\
        v_a^2 =&  2\sum_{k=1}^{\infty}((\lambda+(1-\lambda)\theta^{k}) (1-\lambda+\lambda \theta^{k})^{N-1} -m_a) m_a\nonumber\\
        & + m_a - m_a^2, \\
        v_p^2 = & 2\sum_{k=1}^{\infty}((1-\lambda +\lambda \theta^{k})^{CN} -m_p) m_p +  m_p - m_p^2.
    \end{align*}
    $\hfill\Box$
    \label{thm:mean_vairance}
\end{theorem}

Now, by employing Equations (\ref{eq:active_aoi})--(\ref{eq:passive_moment}), we can represent the moments of AoI for active and passive users in the two-state Markov model as functions of $r$ and $s$. As our next steps, we will determine the optimal $\varsigma=[r,s]$ for the optimization problem given in (\ref{eq:relax_opt_problem}) and then verify the accuracy of our approximations.

\subsection{Optimal System Parameters for Approximated AoI}

In previous sections, we derive approximations for $E[AoI_a^z]$ and $E[AoI_p^z]$ for the two-state Markov model as functions of the probabilities $r$ and $s$. In this section, we further study the optimal behavior of active users and solve for the optimal $r$ and $s$ in the approximated objective function (\ref{eq:relax_opt_problem}). Intriguingly, we discover that when $N>C+4$, the optimal behavior for minimizing $\tilde{F}(\varsigma)$ requires an active user to revert to the Idle state every time a transmission is made, i.e., $s=1$. This result is formally stated in the theorem below.
\begin{theorem}
    When $N>C+4$, there exists a $\lambda\in[0,\frac{1}{N}]$ such that choosing $r=\frac{\lambda}{1-\lambda}$ and $s=1$ minimizes $\tilde{F}([r,s]) = w \sqrt[z]{E[\widetilde{AoI_a}^z]} + (1-w) \sqrt[z]{E[\widetilde{AoI_p}^z]}$.
    \label{theorem:opt_sol}
    $\hfill\Box$
\end{theorem}

This theorem not only reveals an interesting behavior, wherein under the optimal solution, an active user never transmits in two consecutive slots, but also simplifies the search for the optimal solution. It specifically demonstrates that finding the optimal $\lambda$, which can be accomplished through a simple line search, is sufficient for obtaining the optimal solution for minimizing $\tilde{F}([r,s])$.

Prior to proving Theorem~\ref{theorem:opt_sol}, we establish some properties about the optimal $\lambda$ and $\theta$. Recall that $\lambda = \frac{r}{r+s}$ and $\theta=1-r-s$, or, equivalently, $r=\lambda(1-\theta)$ and $s=(1-\lambda)(1-\theta)$. Since $r$ and $s$ are both in the range $[0,1]$, we have $\theta\in[\max\{\frac{\lambda-1}{\lambda},\frac{-\lambda}{1-\lambda}\},1]$.

The proof of Theorem~\ref{theorem:opt_sol} is built upon the following five lemmas.
\begin{lemma}
    For any positive integer $z$, $E[\widetilde{AoI_a}^z]$ is an increasing function of $\frac{1}{m_a}$ and $\frac{v_a^2}{m_a^2}$, and $E[\widetilde{AoI_p}^z]$ is an increasing function of $\frac{1}{m_p}$ and $\frac{v_p^2}{m_p^2}$.
    \label{lemma:incr}
\end{lemma}
\begin{IEEEproof}
    Due to space limitations, we only prove that $E[\widetilde{AoI_a}^z]$ is an increasing function of $\frac{1}{m_a}$ and $\frac{v_a^2}{m_a^2}$.

    Recall the AoI expression for active users in (\ref{eq:active_aoi}), we can observe that each term in (\ref{eq:active_aoi}) is a constant multiplied by $\frac{E[l'_n(i)^{\kappa}]}{E[l'_n(i)]}$, where ${\kappa}\geq 1$. According to (\ref{eq:active_moment}), we have:
    \begin{align}
        \frac{E[l'_n(i)^{\kappa}]}{E[l'_n(i)]} = \sum_{\zeta=0}^{{\kappa}-1}\frac{1}{m_a^{{\kappa}-1}}\frac{({\kappa}-1+\zeta)!}{\zeta!({\kappa}-1-\zeta)!}(\frac{2m_a}{v_a^2})^{-\zeta},
        \label{eq:temp_ratio}
    \end{align}
    whose $\zeta$-th term is $\frac{({\kappa}-1+\zeta)!}{2^{\zeta}\zeta!({\kappa}-1-\zeta)!} \frac{1}{m_a^{{\kappa}-1-\zeta}} \frac{v_a^{2\zeta}}{m_a^{2\zeta}}$. Since ${\kappa}-1\geq \zeta$, this $\zeta$-th term in (\ref{eq:temp_ratio}) is an increasing function of $\frac{1}{m_a}$ and $\frac{v_a^2}{m_a^2}$. Thus, $\frac{E[l'_n(i)^{\kappa}]}{E[l'_n(i)]}$ is an increasing function of $\frac{1}{m_a}$ and $\frac{v_a^2}{m_a^2}$, for any ${\kappa}\geq 1$.

    A similar derivation shows that $E[\widetilde{AoI_p}^z]$ is an increasing function of $\frac{1}{m_p}$ and $\frac{v_p^2}{m_p^2}$.
\end{IEEEproof}

When $\lambda$ is constant, thereby $m_a$ and $m_p$ are fixed, the minimization of $E[\widetilde{AoI_a}^z]$ occurs when $v_a^2$ is at its lowest value, and similarly, $E[\widetilde{AoI_p}^z]$ is minimized when $v_p^2$ is at its smallest. If a specific $\theta$ simultaneously minimizes both $v_a^2$ and $v_p^2$, then it also minimizes the function $\tilde{F}([r,s]) = \tilde{F}([\lambda(1-\theta), (1-\lambda)(1-\theta)])$ under the constraint that $\lambda$ is constant. We denote the optimal solutions for this minimization as $\lambda^*$ and $\theta^*$. In the following three lemmas, we aim to identify the range of values for $\lambda^*$ and $\theta^*$ and determine the optimal value of $\theta^*$.

Based on Lemma~\ref{lemma:incr}, we have the following lemma for $\theta^*$.
\begin{lemma}
    For any $w$, $z$, and $\lambda$, the $\theta$ that minimizes $\tilde{F}([r,s])$ is in range $\theta\leq 0$.
    \label{lemma:theta_r}
\end{lemma}
\begin{IEEEproof}
    We will establish that, for any fixed $\lambda$, $\tilde{F}([r,s])$ increases with $\theta$ when $\theta > 0$. By Lemma 1, we can establish this by showing that $\frac{\partial v_a^2}{\partial \theta} > 0$ and $\frac{\partial v_p^2}{\partial \theta} > 0$ when $\theta > 0$.

    The partial derivative of $v_a^2$ with respect to $\theta$ is:
    \begin{align}
        & \frac{\partial v_a^2}{\partial \theta} = 2m_a\sum_{k=1}^{\infty}[(1-\lambda)k\theta^{k-1}(1-\lambda+\lambda \theta^{k})^{N-1} \nonumber\\
        & \quad +(\lambda+(1-\lambda)\theta^k)(N-1)\lambda k \theta^{k-1}(1-\lambda+\lambda \theta^k)^{N-2})] \nonumber\\
        &=  2m_a \sum_{k=1}^{\infty}[1-2\lambda + N\lambda^2+ N\lambda(1-\lambda)\theta^{k}] \nonumber\\
        & \quad \times k\theta^{k-1}(1-\lambda+\lambda \theta^{k})^{N-2}.
        \label{eq:par_theta}
    \end{align}
    When $\theta>0$, it is easy to verify that each term in (\ref{eq:par_theta}) (i.e., $2m_a$, $[1-2\lambda + N\lambda^2+ N\lambda(1-\lambda)\theta^{k}]$, $\theta^{k-1}$, and $(1-\lambda+\lambda \theta^{k})^{N-2}$) is greater than zero for any $k\in[1,\infty]$. Consequently, the partial derivative of $v_a^2$ with respect to $\theta$ is positive. Similarly, the partial derivative of $v_p^2$ concerning $\theta$ is also positive for $\theta > 0$. 
    
    As a result, the values of $v_a^2$ and $v_p^2$ when $\theta>0$ are higher than their respective values when $\theta = 0$. Therefore, the $\theta$ that minimizes $\tilde{F}([r,s])$ is in range $\theta\leq 0$.
\end{IEEEproof}

Considering the result that $\theta^*\leq 0$, we further derive the following lemma for $\lambda$.
\begin{lemma}
    For any $w$, $z$, and $\theta\leq 0$, the $\lambda$ that minimizes $\tilde{F}([r,s])$ is in the range $[0,\frac{1}{N}]$.
    \label{lemma:opt_l}
\end{lemma}
\begin{IEEEproof}
    See Appendix~\ref{app:opt_l}.
\end{IEEEproof}

Next, we analyze the optimal $\theta$. The lemma below derives the optimal $\theta$ when $\lambda$ is given and fixed.
\begin{lemma}
When $\lambda \in [0, \min\{\alpha,\beta\}]$ is fixed, setting $\theta=-\frac{\lambda}{1-\lambda}$, or, equivalently, setting $r = \frac{\lambda}{1-\lambda}$ and $s = 1$, minimizes $E[\widetilde{AoI_a}^z]$ and $E[\widetilde{AoI_p}^z]$, thereby minimizes $\tilde{F}([r,s])$, where $\alpha$ and $\beta$ are the smallest positive roots of
\begin{align*}
    h_N(y) &=-(N+8)y^3 - (N-13) y^2 - 6y + 1, \\
    \Bar{h}_{CN}(y) &= (CN-10)y^3 - (CN-13)y^2 - 6y + 1,
\end{align*}
respectively.
\label{lemma:opt_theta}
\end{lemma}
\begin{IEEEproof}
    See Appendix~\ref{app:opt_theta}.
\end{IEEEproof}

Comparing Lemma~\ref{lemma:opt_l} and Lemma~\ref{lemma:opt_theta}, we notice that there is a gap between the value range of $\lambda$. Lemma~\ref{lemma:opt_l} works for $\lambda\in[0,\frac{1}{N}]$, while Lemma~\ref{lemma:opt_theta} holds for $\lambda\in[0,\min\{\alpha,\beta\}]$. Therefore, we need to find the relationship between $\alpha$, $\beta$ and $\frac{1}{N}$. In the following lemma, we establish this relationship and show the lower bounds of $\alpha$ and $\beta$ are larger than $\frac{1}{N}$ under some minor conditions.
\begin{lemma} 
    $\alpha > \frac{1}{N}$ when $N > 4$, and $\beta > \frac{1}{N}$ when $N > C+4$.
    \label{lemma_alpha}
\end{lemma}
\begin{IEEEproof}
    See Appendix~\ref{app:alpha}.
\end{IEEEproof}

We are now ready to prove Theorem~\ref{theorem:opt_sol}.
\begin{IEEEproof}[Proof of Theorem~\ref{theorem:opt_sol}]
    It is shown in Lemma~\ref{lemma:opt_l} that the $\lambda^*$ is in the range $[0,\frac{1}{N}]$, which is covered by the range $[0,\min\{\alpha,\beta\}]$ when $N>C+4$ according Lemma~\ref{lemma_alpha}. Therefore, the optimal solution in Lemma~\ref{lemma:opt_theta} holds for $\lambda^*$ when $N>C+4$. Hence, when $N>C+4$, choosing $r=\frac{\lambda^*}{1-\lambda^*}$ and $s=1$ minimizes $E[\widetilde{AoI_a}^z]$ and $E[\widetilde{AoI_p}^z]$, as well as $\tilde{F}([r,s])$.
\end{IEEEproof}

\subsection{Approximation and Optimality Validation}

In this section, we verify whether our results provide a good approximation of any moments of AoI for both active and passive users. Additionally, we evaluate the optimality of the proposed best active user strategy, i.e., changing to the Idle state in the next time slot after each transmission.

For the approximation validation, we consider two different systems. The first is the most basic configuration with a single cluster and one active user, i.e., $C=1$ and $N=1$, while the second system is configured with $C=2$ and $N=4$. For the optimality validation, we examine two other configurations: the first with $C=1$ and $N=6$, and the second with $C=2$ and $N=7$. These configurations differ from those in the approximation validation, as they adhere to the restriction $N>C+4$, as stated in Theorem.~\ref{theorem:opt_sol}.

We note that both $v_a^2$ and $v_p^2$ contain infinite summation terms $\sum_{k=1}^{\infty}$. Since $k$ only appears in $\theta^k$, which converges exponentially to $0$ as $k$ increases, we substitute $\sum_{k=1}^{\infty}$ with $\sum_{k=1}^{1000}$ when computing theoretical $v_n^2$ and $v_p^2$.

For all settings of $N$ and $C$, we simulate various values of $r$, $s$ and $z$. Taking into consideration Lemma~\ref{lemma:opt_l}, we primarily focus on the range where $\frac{r}{r+s}\in(0,\frac{1}{N}]$. In all empirical evaluations, we simulate the system over $100$ distinct runs, each run consisting of $10,000$ time slots, and then compute the average value. 

For approximation validation, we evaluate the mismatches between the theoretical values $\sqrt[z]{E[\widetilde{AoI_a}^z]}$ and $\sqrt[z]{E[\widetilde{AoI_p}^z]}$ and the empirical results $\sqrt[z]{E[AoI_a^z]}$ and $\sqrt[z]{E[AoI_p^z]}$ with different $r$, $s$, and $z$. Mismatches are computed by $\frac{|Theoretical \; value - Empirical \; value|}{Empirical \; value}$. Results are shown in Fig.~\ref{fig:val1} and Fig.~\ref{fig:val2}.

\begin{figure}[!th]
\centering
\captionsetup[subfloat]{labelfont = scriptsize, textfont = scriptsize}
\subfloat[Active user, $s=1$.]{\includegraphics[width=1.72in]{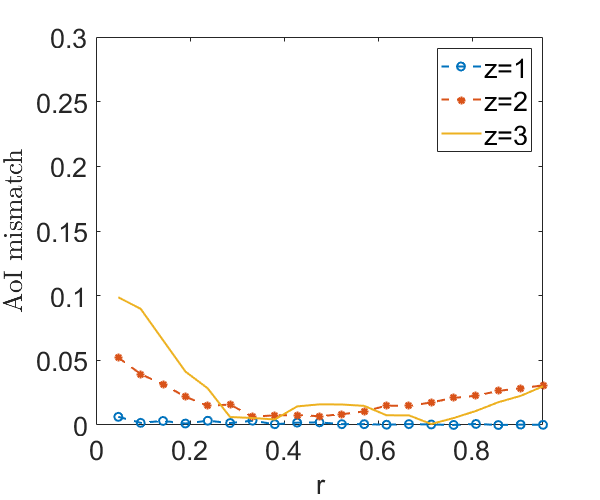}}
\hfil
\subfloat[Passive user, $s=1$.]{\includegraphics[width=1.72in]{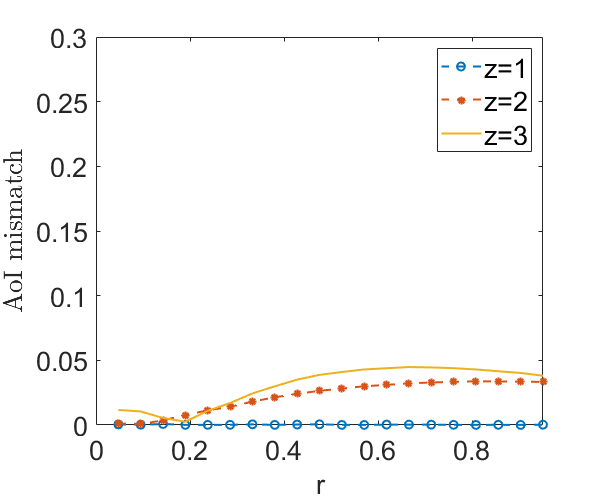}} \\
\hfil
\subfloat[Active user, $s=0.8$.]{\includegraphics[width=1.72in]{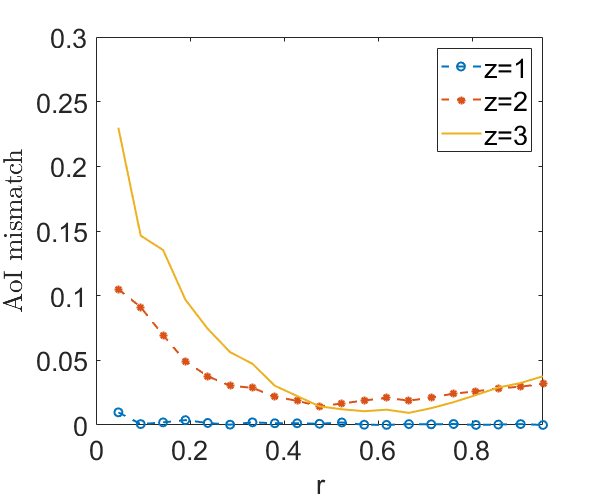}}
\hfil
\subfloat[Passive user, $s=0.8$.]{\includegraphics[width=1.72in]{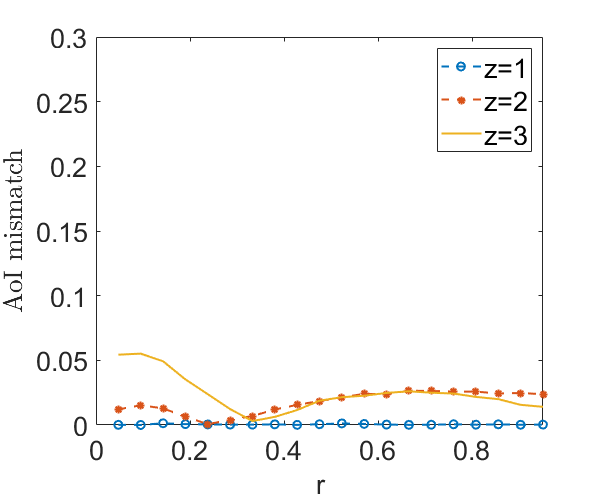}}
\caption{Two-state model approximation mismatches when $C=1$ and $N=1$}
\label{fig:val1}
\end{figure}

\begin{figure}[!th]
\centering
\captionsetup[subfloat]{labelfont = scriptsize, textfont = scriptsize}
\subfloat[Active user, $s=1$.]{\includegraphics[width=1.72in]{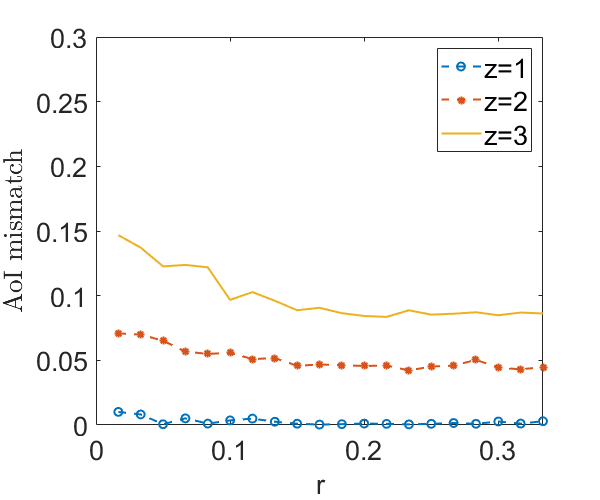}}
\hfil
\subfloat[Passive user, $s=1$.]{\includegraphics[width=1.72in]{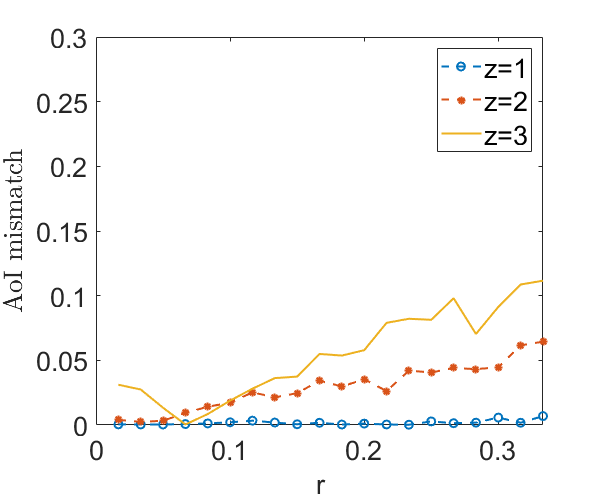}} \\
\hfil
\subfloat[Active user, $s=0.8$.]{\includegraphics[width=1.72in]{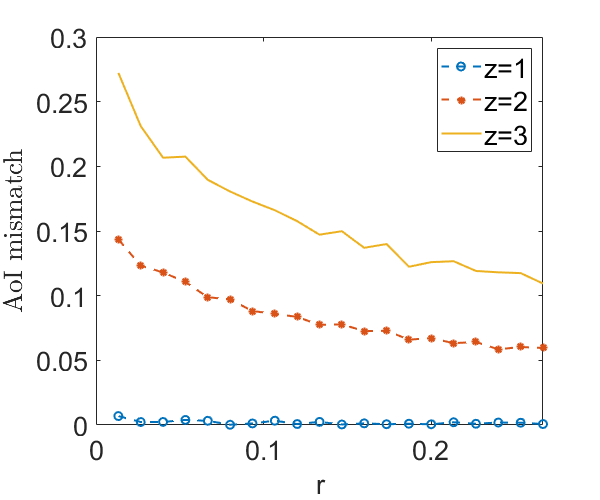}}
\hfil
\subfloat[Passive user, $s=0.8$.]{\includegraphics[width=1.72in]{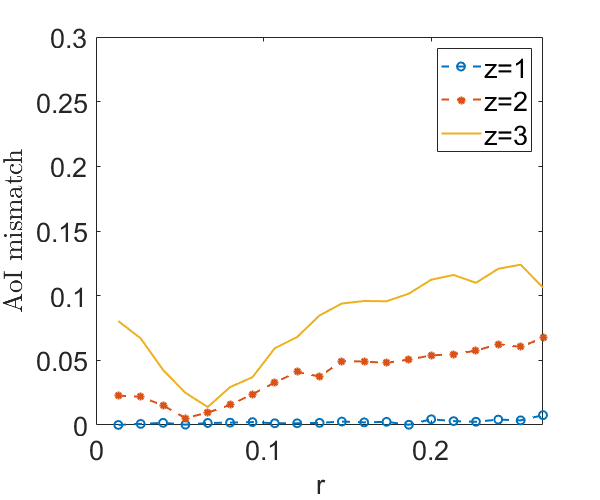}}
\caption{Two-state model approximation mismatches when $C=2$ and $N=4$}
\label{fig:val2}
\end{figure}

From Fig.~\ref{fig:val1} and Fig.~\ref{fig:val2}, we can observe that, when $z=1$, our approximations are highly accurate for both active and passive users, with mismatches being less than $1\%$ in all scenarios. For $z=2$, the mismatches are less than $15\%$ for active users and below $10\%$ for passive users. At $z=3$, active users experience mismatches of less than $25\%$, while passive users see mismatches under $15\%$.

For optimality validation, we assess the differences in empirical values of $\sqrt[z]{E[AoI_a^z]}$ and $\sqrt[z]{E[AoI_p^z]}$ between $s=1$ and the optimal $s$ in practice (with a precision of $0.01$) across various values of $r$ and $z$. These differences are calculated using the formula $\frac{|Value \; for \;S=1 \; - \; Optimal \; value|}{Optimal \; value}$. The results are presented in Fig.~\ref{fig:opt1}.

\begin{figure}[!th]
\centering
\captionsetup[subfloat]{labelfont = scriptsize, textfont = scriptsize}
\subfloat[Active user, $C=1$, $N=6$.]{\includegraphics[width=1.72in]{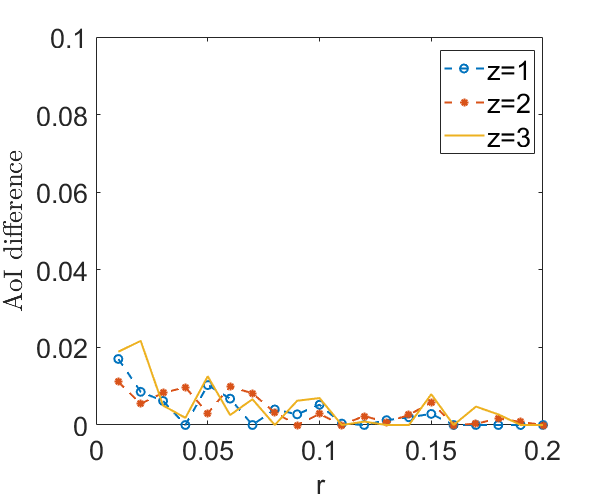}}
\hfil
\subfloat[Passive user, $C=1$, $N=6$.]{\includegraphics[width=1.72in]{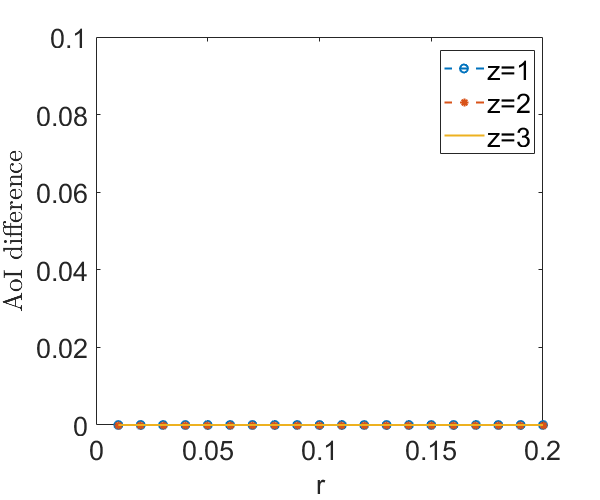}} \\
\subfloat[Active user, $C=2$, $N=7$.]{\includegraphics[width=1.72in]{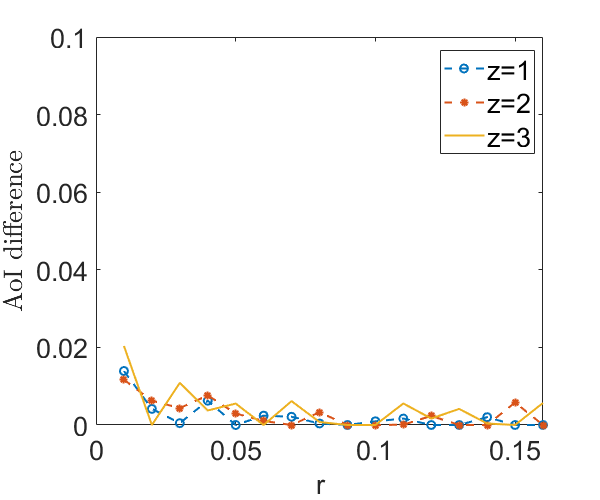}}
\subfloat[Passive user, $C=2$, $N=7$.]{\includegraphics[width=1.72in]{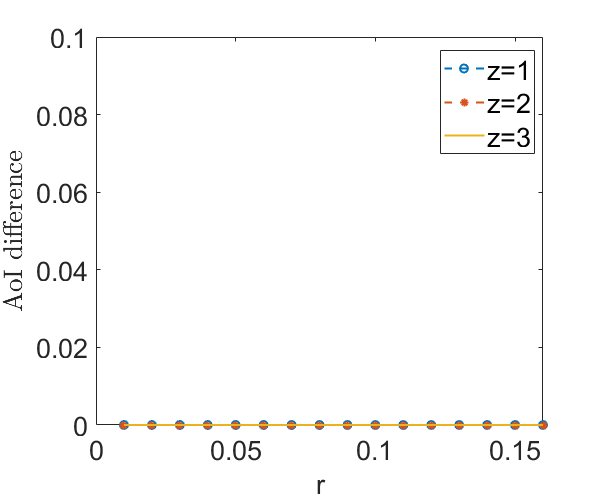}}
\caption{Optimality Validation}
\label{fig:opt1}
\end{figure}

From Fig.~\ref{fig:opt1}, we observe that the proposed strategy $s=1$ is nearly optimal for both active and passive users in all settings for the objective function (\ref{opt_problem}). For active users, the largest observed difference is $2.17\%$ across all settings, while most differences are below $1\%$. In fact, $s=1$ is the optimal setting in $33\%$ cases. For passive users, it is obvious that the optimal setting is $s=1$.

%%%%%%%%%%%%%%%%%%%%%%%%%%%%%%%%%%%%%%%%%%%%%%%%%%%%%%%%%%%%

\section{Wait-and-Go (WaG) Model}
\label{sec:wag}

The optimal solutions of the two-state model for minimizing $\tilde{F}(\varsigma)$ indicate that the best strategy for active users is to remain in the Idle state for at least one time slot after the packet transmission. Inspired by this insight, we introduce a new model for active users. In this model, the active user gains an additional state, referred to as the Wait state, where they stay for $H$ time slots immediately following a transmission.

Fig.~\ref{fig:age_states} illustrates this Markov process. When the active user is in the Idle state at time $t$, it transitions to the TX state at time $t+1$ with a probability $r$. If not, they remain in the Idle state. Once the active user enters the TX state, they transmit a packet in this time slot and will then stay in the Wait state for the next $H$ time slots. After this, the active user returns to the Idle state. This model is referred to as the Wait-and-Go (WaG) model. In this model, $\varsigma$ is the vector $[r,H]$.

\begin{figure}[!th]
\centering
\includegraphics[scale=0.23]{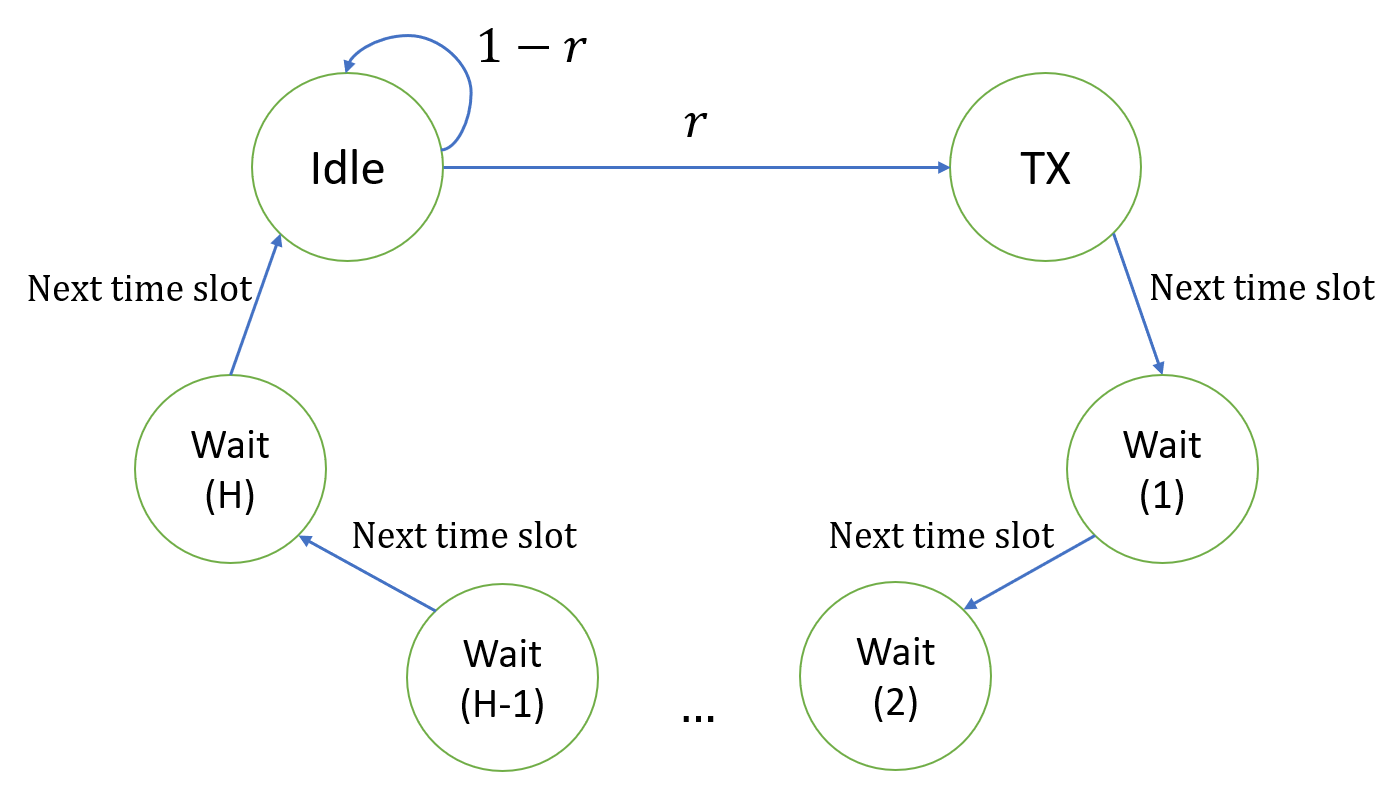}
\caption{Wait-and-Go Model.}
\label{fig:age_states}
\end{figure}

\subsection{Mean and Variance Analysis}

This section develops the means and temporal variances, $m_a$, $m_p$, $v_a^2$, and $v_p^2$, as functions of $r$ and $H$. This allows for the approximation of the moments of AoI of active and passive users as functions of only two parameters, $r$ and $H$. Consequently, the function $\tilde{F}(\varsigma)$ can be expressed as $\tilde{F}([r,H])$.

We first derive the closed-form expressions for means, $m_a$ and $m_p$. Denote $x_{n,t}$ as the state of the active user $n$ at time $t$ and define state $0$ as the Idle state, state $1$ as the TX state, and states $2$ to $H+1$ as the Wait states following the active user being in the TX state (i.e., Wait state $i$ is the state $i-1$ time slots following the TX state). Let $q_i:=Prob(x_{n,t}=i), i\in\{0,1,\dots,H+1\}$ when the system is in a steady state. According to the Markov process shown in Fig.~\ref{fig:age_states}, we have:
\begin{align*}
    q_0 &= (1-r)q_0 + q_{H+1}, \quad q_1 = r q_0, \\
    q_2 &= q_1, \quad q_3=q_2, \quad ..., \quad q_{H+1} = q_{H}.
\end{align*}
Given that $\sum_{i=1}^{H+1}q_i=1$, we have following results:
\begin{align*}
    q_0 = \frac{Hr+1}{(H+1)r+1}, \;  q_1 =q_2=...=q_{H+1}=\frac{r}{(H+1)r+1}.
\end{align*}
Thus, the means can be calculated by:
\begin{align}
    m_a &= q_1 (1-q_1)^{N-1} = \frac{r}{(H+1)r+1}\Big(\frac{Hr+1}{(H+1)r+1}\Big)^{N-1}, \label{eq:mean_active_WaG}\\
    m_p &= (1-q_1)^{CN} = \Big(\frac{Hr+1}{(H+1)r+1}\Big)^{CN}. \label{eq:mean_passive_WaG}
\end{align}

Next, we derive the temporal variances $v_a^2$ and $v_p^2$. Since the derivation techniques for $v_a^2$ and $v_p^2$ are similar, we only demonstrate the derivation of $v_a^2$ for WaG.

Recall the analysis of temporal variances through central limit theorem of the Markov process in Section~\ref{subsec:tx_mv}, we have
\begin{align}
    v_a^2 = Var(S_n(1)) + 2\sum_{k=1}^{\infty} Cov(S_n(1),S_n(k+1)),
    \label{eq:var_iwag}
\end{align}
where $Var(S_n(1))=m_a -m_a^2$, and
\begin{align}
    &Cov(S_n(1), S_n(k+1)) = (Prob(S_n(k+1)=1|S_n(1)=1) \nonumber\\
    & \quad  - m_a)m_a.
\end{align}
As the processes of active users the same, we have the following results:
\begin{align}
    & Prob(S_n(k+1)=1|S_n(1)=1) = Prob(x_{n,k+1}=1|x_{n,1}=1)\nonumber\\
    &  \times \prod_{\zeta \neq n} Prob(x_{\zeta,k+1}\neq 1|x_{\zeta,1}\neq 1), \label{eq:prob_active}
\end{align}
Since $x_{n,k-i}=0$ if $x_{n,k}=i,\forall i\neq 0$, $Prob(x_{n,k}=i|x_{n,1}=j)$ can be expressed as a function of $Prob(x_{n,k-i+j}=0|x_{n,1}=0)$ for any $i$ and $j$ that $k\geq i-j+1$.  For $k < i-j+1$, it is obvious that $Prob(x_{n,k}=i|x_{n,1}=0)=0$. Let $\varphi(k) = Prob(x_{n,k}=0|x_{n,1}=0)$, the next steps are to form $Prob(x_{n,k}=1|x_{n,1}=1)$ and $Prob(x_{n,k}\neq 1|x_{n,1}\neq 1)$ as functions of $\varphi(k)$ and to derive the closed-form of $\varphi(k)$.

Step one, we express $Prob(x_{n,k}=1|x_{n,1}=1)$ in terms of $\varphi(k)$. When $x_{n,1}=1$, it is obvious that we can obtain $x_{n,k+1}=1$ only when $k\geq H+2$, which means
\begin{align*}
    Prob(x_{n,k+1}=1|x_{n,1}=1) = 0, \forall 1\leq k \leq H+1.
\end{align*}
For $k\geq H+2$, we know that if $x_{n,k+1}=1$ and $x_{n,1}=1$, we must have $x_{n,k}=0$ and $x_{n,H+2}=0$. Thus, for $k\geq H+2$,
\begin{align}
    & Prob(x_{n,k+1}=1|x_{n,1}=1) = r Prob(x_{n,k}=0|x_{n,H+2}=0) \nonumber\\
    & = r Prob(x_{n,k-H-1}=0|x_{n,1}=0) = r \varphi(k-H-1).
    \label{eq:prob_11}
\end{align}

Step two, we express $Prob(x_{n,k}\neq 1|x_{n,1}\neq 1)$ as a function of $\varphi(k)$. Specifically, we consider $Prob(x_{n,k+1}=i|x_{n,1}=j)$ where $i,j\neq 1$. According to the Markov process in Fig.~\ref{fig:age_states}, there are two types of $Prob(x_{n,k+1}=i|x_{n,1}=j)$: First, the state will reach state 0 (Idle state) from state $j$, and then reach state $i$. Second, the state will never reach $0$, which only happens when $k=i-j-1$ and $k\leq H+1-j$. For the first case, we have
\begin{align*}
    & Prob(x_{n,k+1}=i|x_{n,1}=j) = r Prob(x_{n,k+1-i}=0|x_{n,1}=j)\\
    & = \begin{cases}
                r Prob(x_{n,k+1-i}=0|x_{n,1}=0), & \text{if $j=0$,} \\
                r Prob(x_{n,k+1-i}=0|x_{n,H+3-j}=0), & \text{otherwise}.
            \end{cases}
\end{align*}
Thus, for the first case, the summation of probabilities is
\begin{align*}
    & \sum_{i\neq 1} \sum_{j\neq 1} Prob(x_{n,k+1}=i|x_{n,1}=j) Prob(x_{n,1}=j|x_{n,1}\neq 1) \\
    & = \sum_{i\neq 1} \big( rProb(x_{n,k+1-i}=0|x_{n,1}=0)Prob(x_{n,1}=0|x_{n,1}\neq 1) \\
    & \quad + \sum_{j=2}^{H+1} rProb(x_{n,k+1-i}=0|x_{n,H+3-j}=0) \\
    & \quad \times Prob(x_{n,1}=j|x_{n,1}\neq 1) \big) \\
    & = \sum_{i\neq 1} \big( \frac{r\varphi(k+1-i)}{Hr+1} + \sum_{j=2}^{H+1} \frac{r^2 \varphi(k-H-1-i+j)}{Hr+1}  \big).
\end{align*}
For the second case, the summation of probabilities is
\begin{align*}
    \sum_{j=2}^{H+1-k}Prob(x_{n,1}=j|x_{n,1}\neq 1) = (H-k) \frac{r}{Hr+1},
\end{align*}
Therefore, we have
\begin{align}
    & Prob(x_{n,k+1}\neq 1|x_{n,1}\neq 1) = \frac{r \max\{H-k,0\}}{Hr+1} \nonumber \\
    & \quad + \sum_{i\neq 1} \big( \frac{r \varphi(k+1-i)}{Hr+1} + \sum_{j=2}^{H+1} \frac{r^2 \varphi(k-H-1-i+j)}{Hr+1}  \big). \label{eq:prob_not11}
\end{align}

In the final step, we derive the closed-form of $\varphi(k)$. Based on Fig.~\ref{fig:age_states}, once $x_{n,t}=1$, $x_{n,t+H+1}=0$ holds. Hence, the state changes of the user $n$ after $x_{n,t}=0$ have only two types: First, the user $n$ will be in the state $0$ in the next time slot, i.e., $x_{n,t+1}=0$. Second, the user $n$ will change to state $1$ in the next time slot, and then go back to state $0$ at time $t+H+2$, i.e., $x_{n,t+H+2}=0$. Therefore, between $x_{n,1}=0$ and $x_{n,k}=0$, there must exist the combination of the first type of state change, which has a length of 1, and the second type of state change, which has a length of $H+2$. Thus, for an integer $\epsilon \in [1,\lfloor \frac{k-1}{H+2} \rfloor]$, the value of $\varphi(k)$, when there are $\epsilon$ second type of state changes between $x_{n,1}=0$ and $x_{n,k}=0$, is:
\begin{align*}
    & \varphi(k)  = \binom{k-1-(H+1)\epsilon}{\epsilon} r^{\epsilon} (1-r)^{k-1-(H+2)\epsilon}.
\end{align*}
If there is no second type of state changes, we have $\varphi(k) = (1-r)^{k-1}$. Thus, we can summarize the following result:
\begin{align}
     \varphi(k)= & \sum_{\epsilon=1}^{\lfloor \frac{k-1}{H+2} \rfloor}\binom{k-1-(H+1)\epsilon}{\epsilon} r^{\epsilon} (1-r)^{k-1-(H+2)\epsilon} \nonumber\\
    &  + (1-r)^{k-1},
    \label{eq:prob0to0}
\end{align}
if $k\geq 1$ and $\varphi(k)=0$ if $k \leq 0$

Combining (\ref{eq:var_iwag})--(\ref{eq:prob0to0}), we have:
\begin{align*}
    & v_a^2 = m_a - m_a^2 + \sum_{k=1}^{\infty}\Big[  r \varphi(k-H-1) \Big ( \frac{r \max\{H-k,0\}}{Hr+1} +\\
    & \sum_{i\neq 1} \big( \frac{r \varphi(k+1-i)}{Hr+1} + \sum_{j=2}^{H+1} \frac{r^2 \varphi(k-H-1-i+j)}{Hr+1}  \big) \Big)^{N-1} \\
    & - m_a \Big]m_a,
\end{align*}
where $i\neq 1$ means $i=0$ or $i\in \{2,3,\dots,H+1\}$, and $\varphi(k)$ is shown in (\ref{eq:prob0to0}).

With a similar derivation, we can obtain expressions for $v_p^2$. In summary, we have the following theorem:
\begin{theorem}
    The means and temporal variances for active and passive users in the WaG are:
    \begin{align*}
        m_a &= q_1 (1-q_1)^{N-1} = \frac{r}{(H+1)r+1}\Big(\frac{Hr+1}{(H+1)r+1}\Big)^{N-1}, \\
        m_p &= (1-q_1)^{CN} = \Big(\frac{Hr+1}{(H+1)r+1}\Big)^{CN},\\
    \end{align*}
    and
    \begin{align*}
        & v_a^2 = m_a - m_a^2 + \sum_{k=1}^{\infty}\Big[  r \varphi(k-H-1) \Big ( \frac{r \max\{H-k,0\}}{Hr+1} + \\
        & \sum_{i\neq 1} \big( \frac{r \varphi(k+1-i)}{Hr+1} + \sum_{j=2}^{H+1} \frac{r^2 \varphi(k-H-1-i+j)}{Hr+1}  \big) \Big)^{N-1} \\
        & - m_a \Big]m_a,
    \end{align*}
    \begin{align*}
        & v_p^2 = m_p - m_p^2 + \sum_{k=1}^{\infty}\Big[ \big ( \frac{r \max\{H-k,0\}}{Hr+1} \\
        & + \sum_{i\neq 1} \big( \frac{r \varphi(k+1-i)}{Hr+1} + \sum_{j=2}^{H+1} \frac{r^2 \varphi(k-H-1-i+j)}{Hr+1}  \big) \Big)^{CN} \\
        & - m_p \Big]m_p,
    \end{align*}
    where
    \begin{align*}
        & \varphi(k) = \sum_{\epsilon=1}^{\lfloor \frac{k-1}{H+2} \rfloor}\binom{k-1-(H+1)\epsilon}{\epsilon} r^{\epsilon} (1-r)^{k-1-(H+2)\epsilon} \\
        & \quad + (1-r)^{k-1}
    \end{align*}
    if $k\geq 1$ and $\varphi(k)=0$ if $k \leq 0$, and $i\neq 1$ means $i=0$ or $i\in \{2,3,\dots,H+1\}$.
    \label{theorem:WaG}
    $\hfill\Box$
\end{theorem}

By combining Theorem~\ref{theorem:aoi_z} and Theorem~\ref{theorem:WaG}, we can approximate the moments of AoI of both active and passive users in WaG in terms of $r$ and $H$. Given that $H$ is an integer and $r$ lies between $0$ and $1$, we can easily determine the optimal settings to solve the optimization problem (\ref{eq:relax_opt_problem}) by implementing grid searches for $r$ and $H$.

\subsection{Approximation Validation and Efficiency Measurement}

In this section, we validate our approximation and measure the efficiency of using theoretical analysis for WaG model.

We validate our approximation through simulations using two different settings: $C=1$ and $N=1$, and $C=2$ and $N=4$. During the simulations, we substitute $\sum_{k=1}^{\infty}$ with $\sum_{k=1}^{1000}$ when computing the theoretical value of $v_n^2$ and $v_p^2$.

We evaluate the mismatches of approximation $\sqrt[z]{E[\widetilde{AoI_a}^z]}$ and $\sqrt[z]{E[\widetilde{AoI_p}^z]}$ by evaluating different combinations of $r$, $H$ and $z$. For all empirical results, we simulate the system over $100$ distinct runs, with each run consisting of $10,000$ time slots, and then compute the average value. Mismatches are computed by $\frac{|Theoretical \; value - Empirical \; value|}{Empirical \; value}$. Results are shown in Fig.~\ref{fig:val3} and Fig.~\ref{fig:val4}.

\begin{figure}[!th]
\centering
\captionsetup[subfloat]{labelfont = scriptsize, textfont = scriptsize}
\subfloat[Active user, $r=0.3$.]{\includegraphics[width=1.72in]{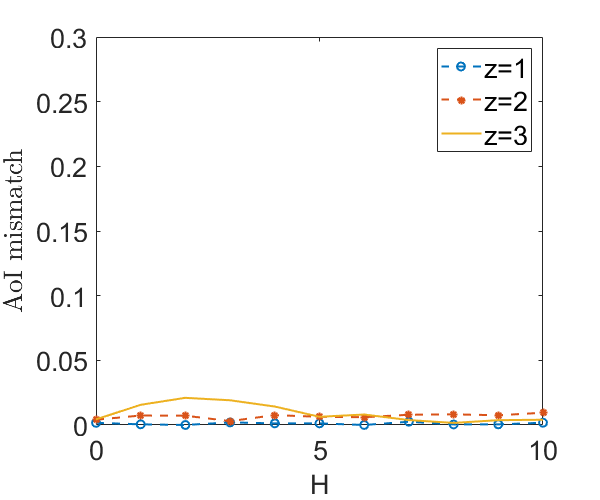}}
\hfil
\subfloat[Passive user, $r=0.3$.]{\includegraphics[width=1.72in]{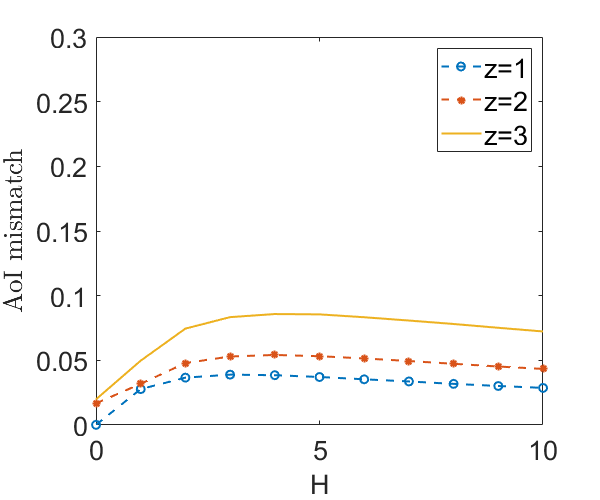}} \\
\subfloat[Active user, $r=0.5$.]{\includegraphics[width=1.72in]{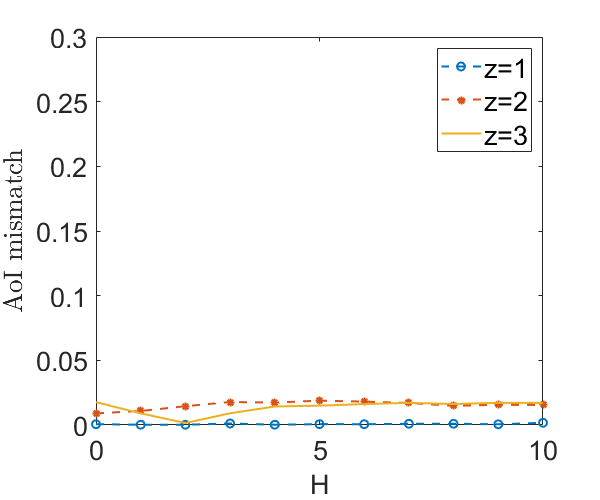}}
\subfloat[Passive user, $r=0.5$.]{\includegraphics[width=1.72in]{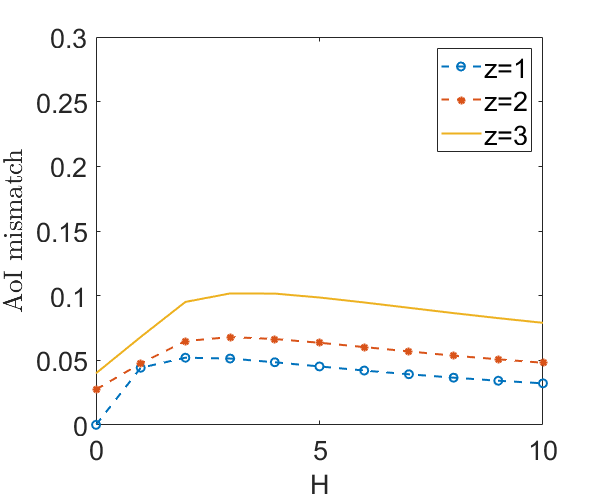}}
\caption{Approximation mismatches for WaG when $C=1$ and $N=1$.}
\label{fig:val3}
\end{figure}

\begin{figure}[!th]
\centering
\captionsetup[subfloat]{labelfont = scriptsize, textfont = scriptsize}
\subfloat[Active user, $r=0.3$.]{\includegraphics[width=1.72in]{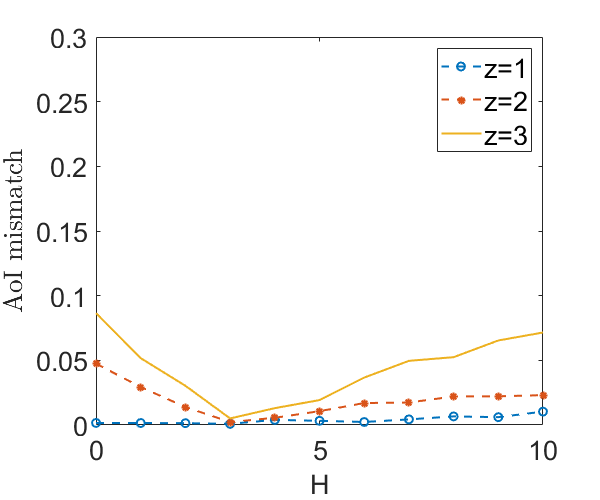}}
\hfil
\subfloat[Passive user, $r=0.3$.]{\includegraphics[width=1.72in]{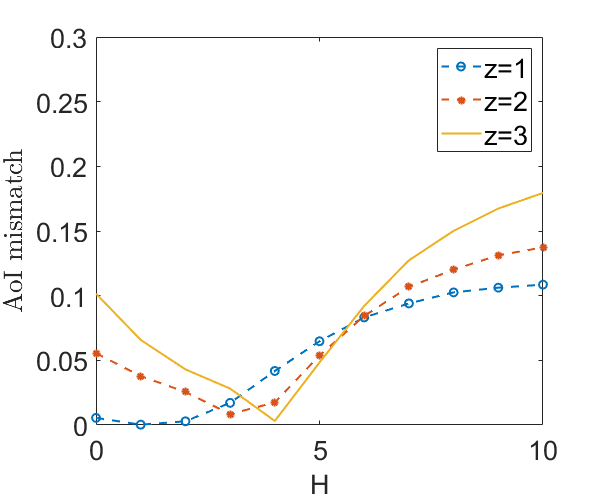}} \\
\hfil
\subfloat[Active user, $r=0.5$.]{\includegraphics[width=1.72in]{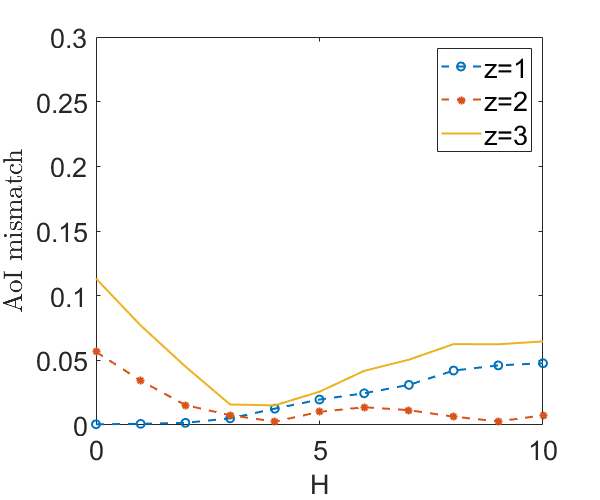}}
\hfil
\subfloat[Passive user, $r=0.5$.]{\includegraphics[width=1.72in]{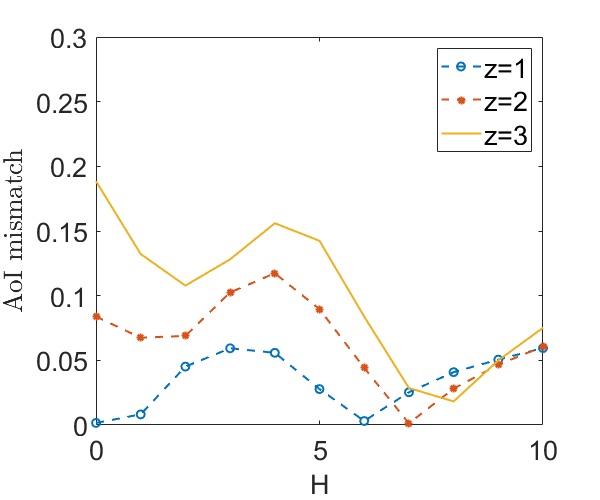}}
\caption{Approximation mismatches for WaG when $C=2$ and $N=4$.}
\label{fig:val4}
\end{figure}

By observing Fig.~\ref{fig:val3} and Fig.~\ref{fig:val4}, we can see that our approximations align closely with the empirical results for active users. For both $z=1$ and $z=2$, the mismatches for active user approximation remain under $5.7\%$. When $z=3$, the mismatches are still limited, falling under $10\%$. For passive users in the system where $c=1$ and $N=1$, the mismatches stay below $10\%$. In contrast, for the system with $C=2$ and $N=4$, they are less than $18\%$.

To assess the efficiency of using theoretical analysis, we consider an alternative parameter-setting method for WaG: WaG with Real-simulation (WaG$_R$). WaG$_R$ adheres to the WaG model and determines the optimal values of $r$ and $H$ based on the average outcomes from $100$  individual simulation runs. We configure the number of time slots in each WaG$_R$ simulation so that the total runtime of WaG$_R$ is the same as our formula-based WaG solution. In subsequent simulations, we conduct an exhaustive search for $r\in(0,1)$ with a precision of $0.01$, and for integer values of $H\in[1,15]$. This search is performed for both WaG$_R$ and the formula-based WaG, aiming to identify the optimal settings for $r$ and $H$.

After the optimal values of $r$ and $H$ are determined, the entire system undergoes 100 individual simulation runs, each spanning 100,000 time slots, to obtain the empirical $F(\varsigma)$. To present the comparison, we calculate the ratio $\frac{F(\varsigma) \; using \; WaG}{F(\varsigma) \; using \; WaG_R}$ for various values of $C$, $N$, $z$, and $w$. The results are displayed in Fig.~\ref{fig:efficy1}.

\begin{figure}[!th]
\centering
\captionsetup[subfloat]{labelfont = scriptsize, textfont = scriptsize}
\subfloat[$C=2$, $w=\frac{N}{N+1}$.]{\includegraphics[width=1.72in]{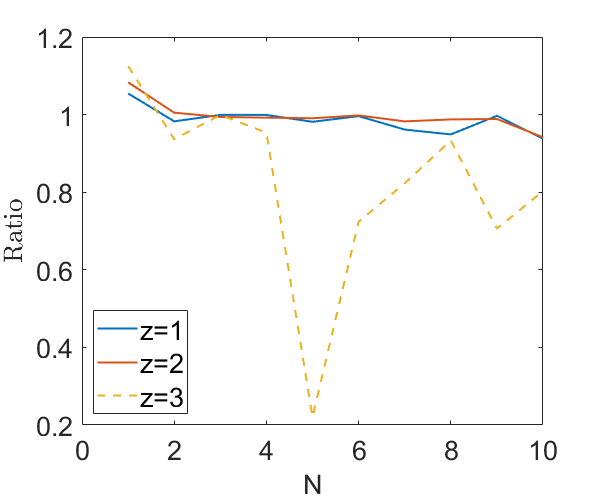}}
\hfil
\subfloat[$C=2$, $w=0.5$.]{\includegraphics[width=1.72in]{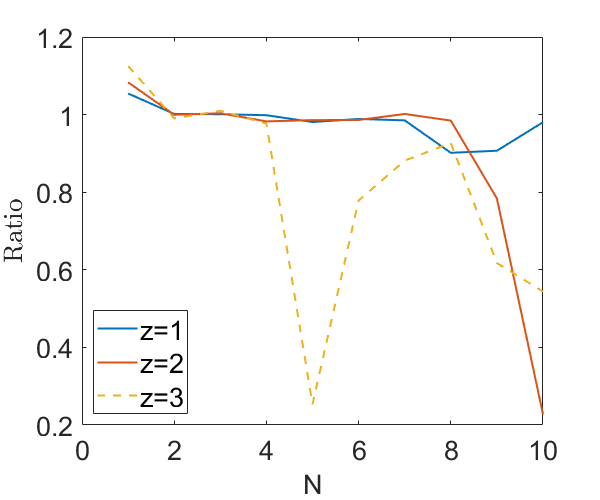}} \\
\hfil
\subfloat[$C=4$, $w=\frac{N}{N+1}$.]{\includegraphics[width=1.72in]{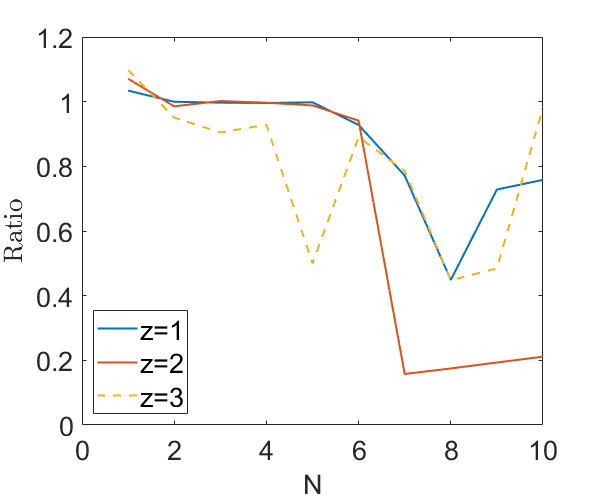}}
\hfil
\subfloat[$C=4$, $w=0.5$. ]{\includegraphics[width=1.72in]{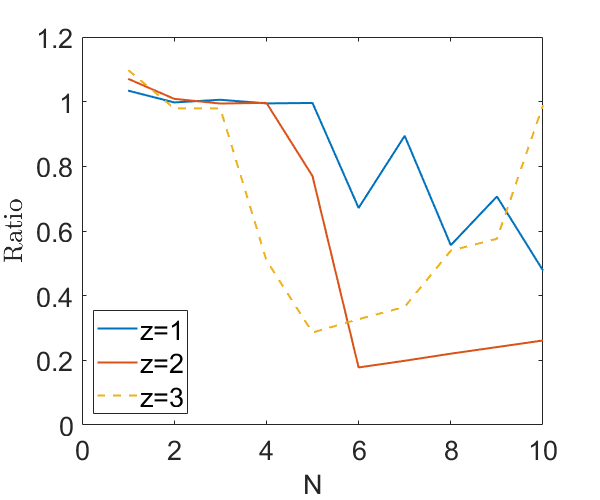}}
\caption{$F(\varsigma)$ comparison between formula-based WaG and simulation based-WaG.}
\label{fig:efficy1}
\end{figure}

As shown in Fig.~\ref{fig:efficy1}, the simulation-based WaG (WaG$_R$) can be more than five times worse than the formula-based WaG. This is because, due to the runtime constraint, the simulation is not long enough to get the accurate value.

%%%%%%%%%%%%%%%%%%%%%%%%%%%%%%%%%%%%%%%%%%%%%%%%%%%%%%%%%%%%%%%%%%%%%%%%%%%%

\section{Simulation Results}
\label{sec:sim}

This section presents simulation results implementing the two models introduced in Sections~\ref{sec:two_states} and ~\ref{sec:wag}. The performance function is the optimization objective in (\ref{opt_problem}), $F(\varsigma)$.

We compare the results from our two-state and WaG models with five other baseline strategies. Below is a detailed description of all seven approaches, along with any necessary modifications made for the simulation settings.
\begin{itemize}
    \item \textbf{Two-State}: This model is shown in Fig.~\ref{fig:two_states} and discussed in Section~\ref{sec:two_states}. Based on Theorem~\ref{theorem:opt_sol}, we choose the optimal $r$ and set $s=1$ for all active users. This optimal $r$ is obtained by plugging $s=1$ into Theorem~\ref{theorem:aoi_z} and Theorem~\ref{thm:mean_vairance}, and doing an exhaustive search over $r$ for $\tilde{F}([r,s])$ with a precision $0.01$.
    \item \textbf{WaG}: This model is shown in Fig.~\ref{fig:age_states} and discussed in Section~\ref{sec:wag}. To find the theoretical optimal $H$ and $r$, we do an exhaustive search over $r\in(0,1)$ with a precision of $0.01$ and over integer $H\in[1,15]$ when applying Theorem~\ref{theorem:WaG}. Then, we use theoretical optimal values of $H$ and $r$ in our simulations.
    \item \textbf{Optimal WaG}: The proposed analysis-based parameter settings for WaG are derived by solving for the optimal solutions of the proposed approximate AoI formula, which may not perfectly match the real optimal parameter settings. Therefore, we conduct an exhaustive search over $r\in(0,1)$ with a precision of $0.01$, and over integer values of $H\in[1,15]$. We then select the best settings after completing the entire simulation and obtaining the final results.
    \item \textbf{Slotted ALOHA}: Within the two-state model, when $r+s=1$, the model simplifies to the slotted ALOHA. This model is renowned and frequently employed for transmissions without acknowledgments. Based on the solutions presented in \cite{yates2017status}, we select $r=\frac{1}{N}$ and $s=1-\frac{1}{N}$ for our simulations.
    \item \textbf{Optimal ALOHA}: After introducing passive users, the optimal $r$ in slotted ALOHA is no longer $\frac{1}{N}$. Thus, we simulate all possible $r$ with precision $0.01$, and then choose the best $r$ to obtain optimal slotted ALOHA. It should be noticed that finding the optimal $r$ may be challenging since it requires simulating the system for a sufficiently long time to determine the optimal $r$ through numeral search.
    \item \textbf{Age Threshold ALOHA (ATA)}: Age threshold ALOHA is proposed by Yavascan and Uysal \cite{yavascan2021analysis}, which is a slotted ALOHA algorithm with an age threshold. If the AoI of one active user is larger than a threshold, this active user will follow slotted ALOHA transmission with a certain probability. Otherwise, this active user will stay silent. However, due to the absence of acknowledgments in our system, the active user cannot be certain of the age of information on the server side. Therefore, in implementing the ATA, we allow the active user to follow the slotted ALOHA protocol once the time elapsed since its last transmission exceeds a specified threshold. As suggested in \cite{yavascan2021analysis}, the transmission probability is set to be $4.69/N$ and the age threshold is $2.2N$.
    \item \textbf{Minimum Hamming User Irrepressible (MHUI)}: The algorithm employs the pre-assigned MHUI sequences as suggested in \cite{liu2023age} for each active user. Once $N$ is finalized, we produce MHUI sequences with the minimal Hamming cross-correlation property based on the CRT construction outlined in Definition 6 of \cite{liu2023age}. These sequences constitute an MHUI sequence set. Each active user within a cluster utilizes one of the sequences to determine the transmission status for every time slot. We then allocate the same MHUI sequence set across different clusters. It's worth noting that this algorithm relies on a central server for sequence assignments, rendering it not a fully distributed algorithm.
\end{itemize}

We conduct simulations with various $C$, $N$ $z$, and $w$. The entire system is simulated across $100$ individual runs, each spanning $100,000$ time slots. The final results are the average of these $100$ runs. We present the value of $F(\varsigma)$ for all five algorithms under each setting. Simulation results are shown in Fig.~\ref{fig:sim1} and Fig.~\ref{fig:sim2}. Additionally, we conduct a runtime comparison between the proposed WaG and the optimal WaG. This comparison is measured by the formula: $\frac{\textit{runtime of Optimal WaG}}{\textit{runtime of WaG}}$, and is presented in Fig.~\ref{fig:sim3}.

\begin{figure}[!th]
\centering
\captionsetup[subfloat]{labelfont = scriptsize, textfont = scriptsize}
\subfloat[$N=8$, $w=\frac{N}{N+1}$.]{\includegraphics[width=1.72in]{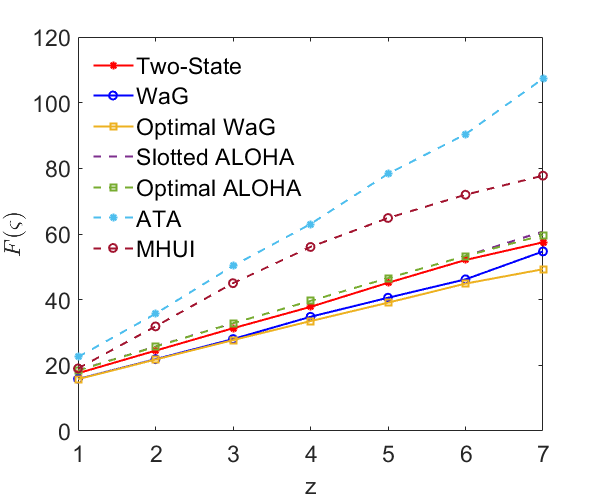}}
\hfil
\subfloat[$N=8$, $w=0.5$.]{\includegraphics[width=1.72in]{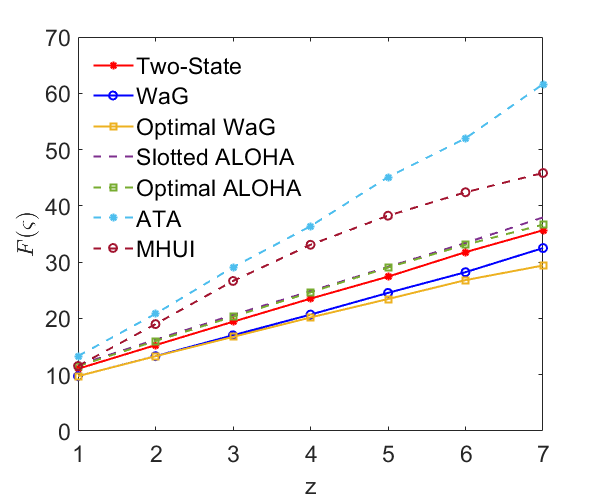}} \\
\hfil
\subfloat[$N=10$, $w=\frac{N}{N+1}$.]{\includegraphics[width=1.72in]{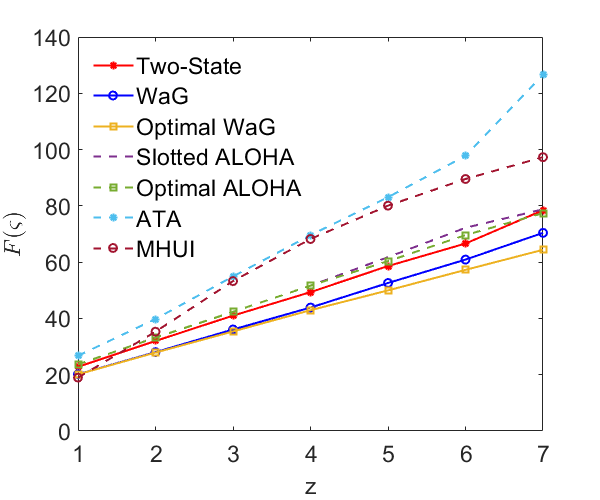}}
\hfil
\subfloat[$N=10$, $w=0.5$.]{\includegraphics[width=1.72in]{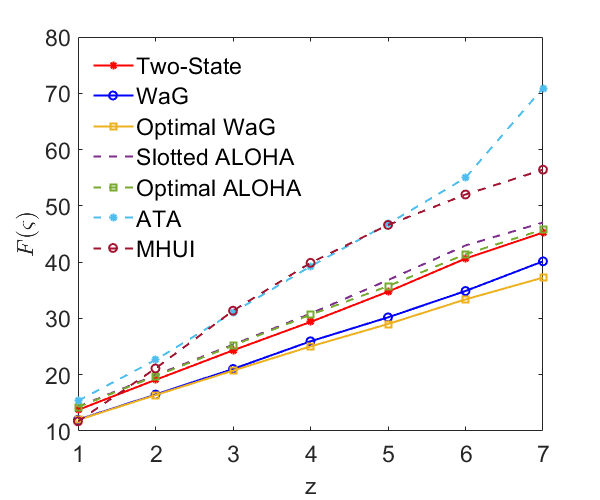}}
\caption{$F(\varsigma)$ when $C=1$.}
\label{fig:sim1}
\end{figure}

\begin{figure}[!th]
\centering
\captionsetup[subfloat]{labelfont = scriptsize, textfont = scriptsize}
\subfloat[$N=8$, $w=\frac{N}{N+1}$.]{\includegraphics[width=1.72in]{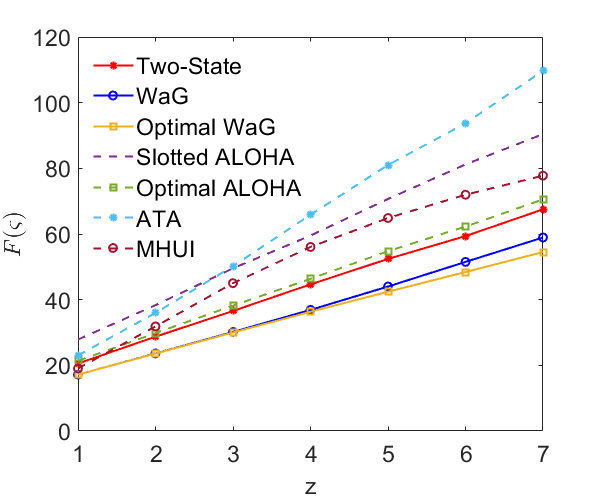}}
\hfil
\subfloat[$N=8$, $w=0.5$.]{\includegraphics[width=1.72in]{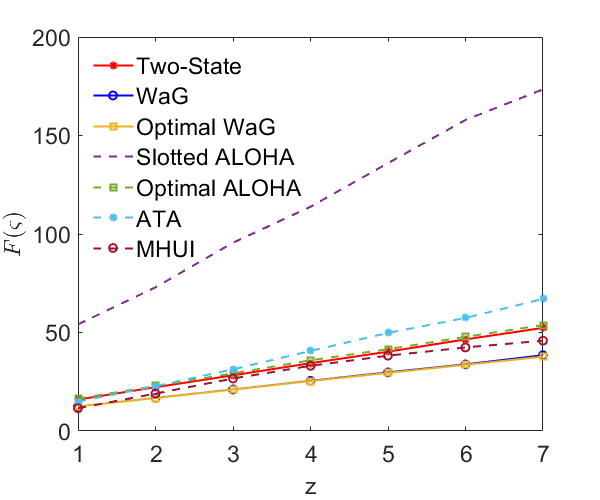}} \\
\hfil
\subfloat[$N=10$, $w=\frac{N}{N+1}$.]{\includegraphics[width=1.72in]{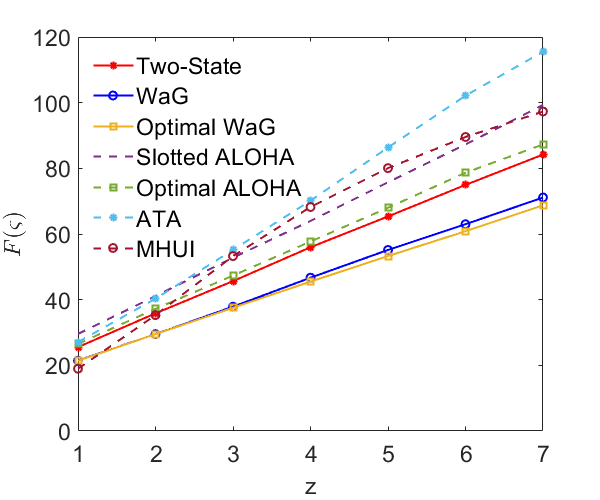}}
\hfil
\subfloat[$N=10$, $w=0.5$.]{\includegraphics[width=1.72in]{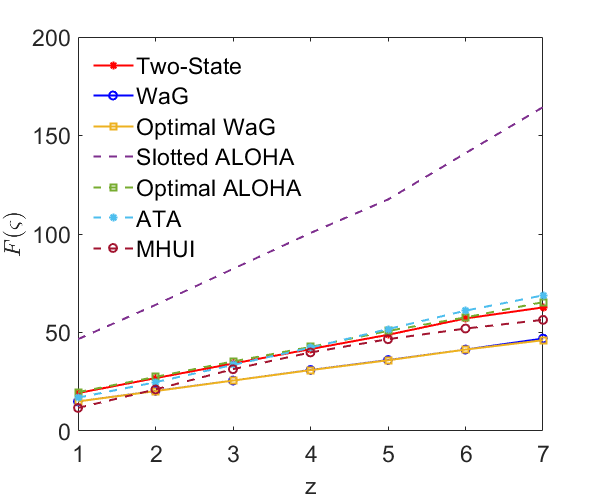}}
\caption{$F(\varsigma)$ when $C=4$.}
\label{fig:sim2}
\end{figure}

\begin{figure}[!th]
\centering
\captionsetup[subfloat]{labelfont = scriptsize, textfont = scriptsize}
\subfloat[$C=1$.]{\includegraphics[width=1.72in]{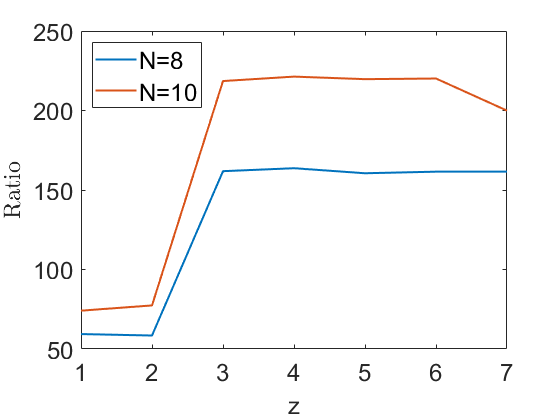}}
\hfil
\subfloat[$C=4$.]{\includegraphics[width=1.72in]{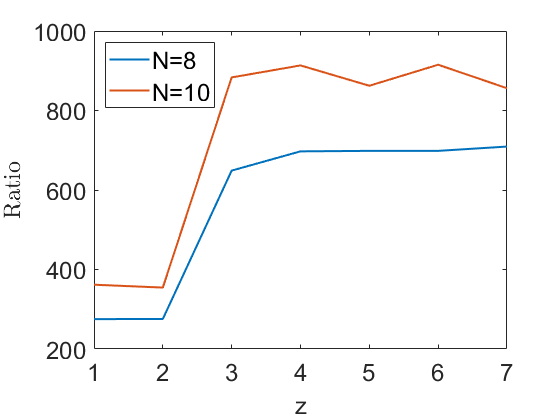}}
\caption{Runtime Comparison between WaG and optimal WaG.}
\label{fig:sim3}
\end{figure}

From both Fig.~\ref{fig:sim1} and Fig.~\ref{fig:sim2}, it is clear that our two models consistently outperform the three ALOHA-based baseline strategies across various settings. There are several reasons why ATA performs poorly in simulations. Firstly, ATA is focused solely on optimizing the AoI for active users. Secondly, the parameter settings in \cite{yavascan2021analysis} are based on the assumption of a significantly large number of users. Moreover, ATA is originally designed to rely on acknowledgments, which are not available in our system. Consequently, the performance of ATA can be very poor, even when $z=1$. However, we have yet to identify a more suitable strategy that can be implemented in our system.

Moreover, our WaG model surpasses the MHUI in nearly all scenarios, with the only exception of when $z=1$ and $N=10$. When examining WaG in detail, we find that the improvements achieved by employing WaG, when compared to the best-performing among the four state-of-the-art strategies, can exceed an average of $15\%$ reduction in any given setting of $C$, $N$ and $w$.

Furthermore, the comparison of $F(\varsigma)$ between WaG and its optimal performance demonstrates that our proposed algorithm is indeed near-optimal. For $C=1$, our proposed theoretical WaG method exhibits an $F(\varsigma)$ that is at most $11\%$ higher, and for $C=4$, it is at most $8\%$ higher across all settings. It is also evident that when $z$ is small, specifically $z=1$ and $z=2$, the proposed method is almost optimal. Additionally, we observe that as $N$ and $C$ increase, the performance of the analysis-based parameter settings aligns more closely with the optimal parameter settings. Moreover, the comparison shown in Fig.~\ref{fig:sim3} demonstrates that the runtime of the optimal WaG can be hundreds of times higher than that of WaG. This significant difference in runtime demonstrates the efficiency and utility of our solution.

%%%%%%%%%%%%%%%%%%%%%%%%%%%%%%%%%%%%%%%%%%%%%%%%%%%%%%%%%%%%%%%%%%%%%%%%%%%%%

\section{Future Work}
\label{sec:future}
In this section, we describe several limitations of this work and propose future work to overcome these limitations. Firstly, our approach requires formulating the mean and temporal variance as functions of Markovian process parameters. Obtaining a closed form for these functions can be challenging, especially for complex protocols. In future research, it would be worthwhile to explore whether there is a uniform approach that can easily formulate the mean and temporal variance and be applicable to all Markov processes. For example, it would be interesting to study if there exists a method that calculates the mean and temporal variance based on the transition matrix. Secondly, our system model is currently limited to symmetric cases. In future studies, it would be interesting to expand our analysis to include asymmetric cases, such as those involving heterogeneous passive users and asymmetric groups of active users. Thirdly, in this paper, we examine the tightness of the proposed approximation method through simulations. It would also be interesting to establish an approximation bound for our approach.

%%%%%%%%%%%%%%%%%%%%%%%%%%%%%%%%%%%%%%%%%%%%%%%%%%%%%%%%%%%%%%%%%%%%%%%%%%%%

\section{Conclusion}
\label{sec:conclusion}

This paper delves into the optimization of the moments of Age of Information (AoI) for both active and passive users within a random access network. In this setup, active users transmit their data following a Markov process, which can lead to mutual interference. Leveraging a second-order model, we approximate the moments of AoI as functions of both the mean and temporal variance of successful transmissions. We then analyze a basic two-state Markov model, expressing the approximated AoI in terms of its model parameters. This allows us to optimize the moments of AoI for both active and passive users. Intriguingly, the optimal strategy for the approximated objective function suggests that the most effective strategy for active users within this model is to become silent immediately after one transmission. The simulation results verified that this strategy is optimal in $33\%$ of cases and near optimal in other. Inspired by this insight, we introduce an alternative model, termed Wait-and-Go (WaG), and approximate its AoI based on two specific model parameters. Empirical validations confirm that our moments of AoI approximations are considerably accurate for both models. Through simulations, we demonstrate that our WaG model, based on theoretical parameter settings, outperforms four baseline algorithms, including a recent proposed algorithm, in nearly all settings. The simulations also indicate that the proposed theoretical-based parameter settings are near optimal in all settings.

%%%%%%%%%%%%%%%%%%%%%%%%%%%%%%%%%%%%%%%%%%%%%%%%%%%%%%%%%%%%%%%%%%%%%%%%%%%%%%

\section*{Acknowledgments}
This material is based upon work supported in part by NSF under Award Number ECCS-2127721 and in part by the U.S. Army Research Laboratory and the U.S. Army Research Office under Grant Number W911NF-22-1-0151.

\appendices

\section{Proof of Lemma \ref{lemma:opt_l}}
\label{app:opt_l}
    We first establish the following claim: $E[\widetilde{AoI_p}^z]$ is an increasing function of $\lambda$ for a fixed $\theta \leq 0$. By lemma~\ref{lemma:incr}, we can establish this claim by showing that both $\frac{1}{m_p}$ and $\frac{v_p^2}{m_p^2}$ increases with $\lambda$ for any $\theta$ that $\theta \leq 0$.

    Recall $m_p = (1-\lambda)^{CN}$ and $v_p^2 = 2\sum_{k=1}^{\infty}((1-\lambda +\lambda \theta^{k})^{CN} -m_p) m_p +  m_p - m_p^2$.

    It is obvious that $m_p$ decreases, thereby $\frac{1}{m_p}$ increases, with $\lambda$.

    For $\frac{v_p^2}{m_p^2}$, do the partial derivative with respect to $\lambda$ and we have
    \begin{align*}
         & \frac{\partial (v_p^2/m_p^2)}{\partial \lambda} = 2\sum_{k=1}^{\infty} \Big[ \frac{CN(\theta^k-1)(1-\lambda+\lambda\theta^k)^{CN-1}m_p}{m_p^2} \\
         & + \frac{CN(1-\lambda)^{CN-1}(1-\lambda+\lambda\theta^k)^{CN}}{m_p^2} \Big] + \frac{CN(1-\lambda)^{CN}}{m_p^2} \\
         %& = \sum_{k=1}^{\infty}\frac{2CN(1-\lambda)^{CN-1}(1-\lambda+\lambda\theta^k)^{CN-1}}{m_p^2} [(\theta^k-1)(1-\lambda) \\
         %& +1-\lambda+\lambda\theta^k ] + \frac{CN(1-\lambda)^{CN}}{m_p^2} \\
         & = \frac{2CN(1-\lambda)^{CN-1}[1-\lambda + \sum_{k=1}^{\infty}(1-\lambda+\lambda\theta^k)^{CN-1}\theta^k]}{m_p^2}.
    \end{align*}
    Since $\theta \in (-1,0)$ and $1-\lambda+\lambda\theta^k \geq 0$ for all $k$,
    \begin{align*}
         & \frac{\partial (v_p^2/m_p^2)}{\partial \lambda} \\
         & \geq \frac{2CN(1-\lambda)^{CN-1}[1-\lambda + \sum_{k=1}^{\infty}(1-\lambda)^{CN-1}\theta^k]}{m_p^2} \\
         & = \frac{CN(1-\lambda)^{CN-1}[1-\lambda + (1-\lambda)^{CN-1} \frac{\theta}{1-\theta}]}{m_p^2}.
    \end{align*}
    When $\lambda\in [0,\frac{1}{2}]$, $\theta\in [-\frac{\lambda}{1-\lambda},0]$, and hence
    \begin{align*}
        & \frac{\partial (v_p^2/m_p^2)}{\partial \lambda} \geq \frac{2CN(1-\lambda)^{CN-1}[1-\lambda - (1-\lambda)^{CN-1} \lambda]}{m_p^2} \\
        & = \frac{2CN(1-\lambda)^{CN-1}[1-(1+(1-\lambda)^{CN-1})\lambda]}{m_p^2} \\
        & \geq \frac{2CN(1-\lambda)^{CN-1}[1-(1+1)\frac{1}{2}]}{m_p^2} = 0.
    \end{align*}
    When $\lambda\in [\frac{1}{2},1]$, $\theta\in [\frac{\lambda-1}{\lambda}, 0]$, and then
    \begin{align*}
        & \frac{\partial (v_p^2/m_p^2)}{\partial \lambda} \geq \frac{2CN(1-\lambda)^{CN-1}[1-\lambda - (1-\lambda)^{CN}]}{m_p^2} \\
        & \geq \frac{2CN(1-\lambda)^{CN-1}[1-\lambda - (1-\lambda)]}{m_p^2} = 0.
    \end{align*}
    Therefore, $\frac{\partial (v_p^2/m_p^2)}{\partial \lambda} \geq 0$.

    So, both $\frac{1}{m_p}$ and $\frac{v_p^2}{m_p^2}$ increases with $\lambda$ for any $\theta$ that $\theta \leq 0$.

    A similar but more involved analysis can establishes the following claim: $E[\widetilde{AoI_a}^z]$ is an increasing function when $\lambda>\frac{1}{N}$ and $\theta \leq 0$.

    Therefore, $\tilde{F}([r,s]) = w \sqrt[z]{E[\widetilde{AoI_a}^z]} + (1-w) \sqrt[z]{E[\widetilde{AoI_p}^z]}$ is an increasing function of $\lambda$ when $\lambda>\frac{1}{N}$, for any $\theta \leq 0$. Hence, the optimal $\lambda$ that minimizes $\tilde{F}([r,s])$ is in the range $[0,\frac{1}{N}]$.

\section{Proof of Lemma \ref{lemma:opt_theta}}
\label{app:opt_theta}
    When $\lambda$ is fixed, $m_a = \lambda (1-\lambda)^{N-1}$ and $m_p=(1-\lambda)^{CN}$ are determined. Hence, by Theorem~\ref{theorem:aoi_z}, $E[\widetilde{AoI_a}^z]$ and $E[\widetilde{AoI_p}^z]$ are increasing functions of $v_a^2$ and $v_p^2$ respectively. By showing that the minimum $\theta$ minimizes $v_a^2$ and $v_p^2$ when $\lambda \leq \min\{\alpha,\beta\}$, we can then state that the minimum $\theta$, which is $-\frac{\lambda}{1-\lambda}$, minimizes $E[\widetilde{AoI_a}^z]$ and $E[\widetilde{AoI_p}^z]$.

    Due to space limitation, we only show the details of the proof of results for passive user, i.e., the minimum $\theta$ minimizes $v_p^2$ when $\lambda \leq \beta$. This is established by showing $\frac{\partial v_p^2}{\partial \theta} \geq 0$ for $\lambda \leq \beta$.
    
    Recall the expression of $v_p^2$, 
    \begin{align*}
        v_p^2 &= 2\sum_{k=1}^{\infty}((1-\lambda +\lambda \theta^{k})^{CN} -m_p) m_p +  m_p - m_p^2.
    \end{align*}
    It is easy to verify that $v_p^2\at{\theta>0} \geq v_p^2\at{\theta=0}$. Therefore, $v_p^2$ reaches its minimum in the region $-\frac{\lambda}{1-\lambda} \leq \theta \leq 0$. The partial derivative of $v_p^2$ with respect to $\theta$ is
    \begin{align}
        \frac{\partial v_p^2}{\partial \theta} = 2m_p \sum_{k=1}^{\infty}CN k \lambda \theta^{k-1} (1 - \lambda + \lambda \theta^k)^{CN-1}.
        \label{eq:theta_par}
    \end{align}
    Let $\phi(k)=2 m_p CN k \lambda \theta^{k-1} (1 - \lambda + \lambda \theta^k)^{CN-1}$ be the $k-th$ term in (\ref{eq:theta_par}). When $\theta \in [-\frac{\lambda}{1-\lambda},0]$, $\phi(k) \geq 0$ when $k$ is odd and $\phi(k) < 0$ when $k$ is even. 
    
    Thus, $\frac{\partial v_p^2}{\partial \theta}\geq 0$ if $-\frac{\phi(2i-1)}{\phi(2i)} \geq 1$ for all positive integer $i$. Next, we show that $-\frac{\phi(2i-1)}{\phi(2i)} \geq 1$ holds when  $\lambda \leq \beta$. Plugging in the formula of $\phi(k)$, we have
    \begin{align*}
        -\frac{\phi(2i-1)}{\phi(2i)} = - \frac{2i-1}{2i}\frac{1}{\theta} \Big( \frac{1 - \lambda + \lambda \theta^{2i-1}}{1 - \lambda + \lambda \theta^{2i}} \Big)^{CN-1}.
    \end{align*}
    Since $\theta^{2i-1}\geq \theta \geq -\frac{\lambda}{1-\lambda}$ and $\theta^{2i}\leq \theta^{2} \leq \frac{\lambda^2}{(1-\lambda)^2}$ when $\theta \in [-\frac{\lambda}{1-\lambda},0]$, we further have
    \begin{align*}
        -\frac{\phi(2i-1)}{\phi(2i)} &\geq -\frac{\phi(1)}{\phi(2)} = -\frac{1}{2} \frac{1}{\theta} \Big( \frac{1-\lambda+\lambda \theta}{1-\lambda+\lambda \theta^{2}} \Big)^{CN-1} \\
        %&\geq \frac{1-\lambda}{2\lambda} \Big( \frac{2\lambda^2-3\lambda+1}{3\lambda^2-3\lambda+1} \Big)^{CN-1} \\
        & \geq \frac{1-\lambda}{2\lambda} \Big( 1 - \frac{\lambda^2}{3\lambda^2-3\lambda+1} \Big)^{CN-1}.
    \end{align*}
    To simplify the expressions, we define $A_{CN}(y)$ and $B(y)$ as follows.
    \begin{align*}
        A_{CN}(y) := \Big( 1 - \frac{y^2}{3y^2-3y+1} \Big)^{CN-1}, \quad B(y) := \frac{2y}{1-y}.
    \end{align*}
    Hence, $-\frac{\phi(2i-1)}{\phi(2i)} \geq \frac{A_{CN}(\lambda)}{B(\lambda)}$. When $y\in[0, \frac{1}{2}]$, we have
    \begin{align*}
        &A_{CN}(y) - B(y) = \Big( 1 - \frac{y^2}{3y^2-3y+1} \Big)^{CN-1} - \frac{2y}{1-y} \\
        & \geq 1- (CN-1)\frac{y^2}{3y^2-3y+1} - \frac{2y}{1-y} \\
        %& = \frac{(CN-10)y^3-(CN-13)y^2-6y+1}{(3y^2-3y+1)(1-y)} \\
        & = \frac{\Bar{h}_{CN}(y)}{(3y^2-3y+1)(1-y)}.
    \end{align*}
    Note $\Bar{h}_{CN}(0) = 1$, $\Bar{h}_{CN}(\frac{1}{2}) = -\frac{CN}{8}$, and $\beta$ is the smallest positive root of $\Bar{h}_{CN}(y)$. Thus, $\Bar{h}_{CN}(y)\geq 0$ if $y\in[0,\beta]$ and $\beta < \frac{1}{2}$. So, when $y\in[0,\beta]$, $3y^2-3y+1 > 0$ and $1-y>0$, and hence
    \begin{align*}
        A_{CN}(y) - B(y) = \frac{\Bar{h}_{CN}(y)}{(3y^2-3y+1)(1-y)} \geq 0.
    \end{align*}
    Therefore, if $\lambda \in [0,\beta]$, $-\frac{\phi(2i-1)}{\phi(2i)} \geq \frac{A_{CN}(\lambda)}{B(\lambda)} \geq 1$.

    So, $\frac{\partial v_p^2}{\partial \theta} > 0$ and $v_p^2$ achieves its minimum value when $\theta$ is minimized and $\lambda \in [0,\beta]$. Thus, choosing $r=\frac{\lambda}{1-\lambda}$ and $s=1$ minimizes $v_p^2$. 

    A similar but more complicated derivation can show that the minimum $\theta$ minimizes $v_a^2$ when $\lambda\leq \alpha$. %Details can be found in the Appendix C of our arXiv paper \cite{fan2023minimizing}.

    Therefore, when $\lambda\in \min\{\alpha,\beta\}$, $\theta=-\frac{\lambda}{1-\lambda}$, which is achieved by setting $r=\frac{\lambda}{1-\lambda}$ and $s=1$, minimizes $\tilde{F}([r,s])$.

\section{Proof of Lemma \ref{lemma_alpha}}
\label{app:alpha}
    To prove the lemma, we will establish the following claims: First, both $h_N(y)$ and $\Bar{h}_{CN}(y)$ have only one root in the interval $[0,1]$. Second, $h_N(y) > 0$ for $y=0$ and $y=\frac{1}{N}$ when $N>4$. Finally, $\Bar{h}_{CN}(y) > 0$ for $y=0$ and $y=\frac{1}{N}$ when $N>C+4$. Based on these claims and considering $\alpha$ is the smallest positive root of $h_N(y)$, $\alpha$ cannot be in the interval $[0,\frac{1}{N}]$, i.e., $\alpha>\frac{1}{N}$, when $N>4$. Similarly, we can conclude that $\beta>\frac{1}{N}$ when $N > C+4$.

    In the following, we first show that $h_N(y)$ has only one root in $[0,1]$. Let us rewrite the expression of $h_N(y)$.
    \begin{align*}
        h_N(y) = -(N+8)y^3 - (N-13) y^2 - 6y + 1.
    \end{align*}
    The derivative of $h_N(y)$ is
    \begin{align*}
        h'_N(y) = -3(N+8)y^2-2(N-13)y-6.
    \end{align*}
    Let $h'_N(y) = 0$, we have $y_1=\frac{2(N-13) + \sqrt{4(N-13)^2-72(N+8)}}{-6(N+8)}$ and $y_2=\frac{2(N-13) - \sqrt{4(N-13)^2-72(N+8)}}{-6(N+8)}$. Thus, the roots of $h'_N(y)$ are either not existing or all negative. Since there is no root when $y \geq 0$ and $h'_N(0)=-6<0$, $h'_N(y)$ is always negative when $y\geq 0$. Therefore, $h_N(y)$ is always decreasing when $y\geq 0$. Considering $h_N(0)=1>0$ and $h_N(1)=-2N<0$, $\alpha$ must exist between $0$ and $1$ and there is only one root for $h_N(y)$ when $y\in[0,1]$. Besides, $h_N(y)>0$ if and only if $y<\alpha$, when $y \geq 0$.

    Now, we prove that $h_N(y) > 0$ for $y=0$ and $y=\frac{1}{N}$ when $N>4$. Set $y=0$, we have $h_N(0) = 1>0$. Let $y=\frac{1}{N}$, we have $h_N(\frac{1}{N}) = \frac{N^3-7N^2+12N-8}{N^3}$. Denote $\varepsilon(N):=N^3-7N^2+12N-8$. Set $\varepsilon(N)=0$, we obtain the only real root, which is approximately $4.8751$. Since $\varepsilon(N)\rightarrow \infty$ when $N\rightarrow \infty$ and $N$ is an integer, we have $\varepsilon(N)>0$ when $N > 4$. Thus, $h_N(\frac{1}{N})=\frac{\varepsilon(N)}{N^3}>0$ when $N>4$.

    Therefore, the two claims for $\alpha$ and $h_N(y)$ have been proved. So, $\alpha>\frac{1}{N}$, when $N>4$.

    Next, we show that $\Bar{h}_{CN}(y)$ has only one root in $[0,1]$. Let us rewrite the expression of $\Bar{h}_{CN}(y)$.
    \begin{align*}
        \Bar{h}_{CN}(y) = (CN-10)y^3-(CN-13) y^2 -6y +1.
    \end{align*}
    The rest of analysis of $\Bar{h}_{CN}(y)$ depends on the sign of $CN-10$.
    \begin{itemize}
        \item When $CN-10=0$, $\Bar{h}_{CN}(y)=3y^2-6y+1$. There are two roots for $\Bar{h}_{CN}(y)$, which are $\frac{3-\sqrt{6}}{3}\approx 0.1835$ and $\frac{3+\sqrt{6}}{3} \approx 1.8165$. In this case, only one root exists in $[0,1]$.
        \item When $CN-10 > 0$, $\Bar{h}_{CN}(-\infty) < 0$, $\Bar{h}_{CN}(0)=1>0$, $\Bar{h}_{CN}(1)=-2<0$, and $\Bar{h}_{CN}(\infty) > 0$. Thus, $\Bar{h}_{CN}(y)$ must have three roots, one is less than $0$, one is between $0$ and $1$, and one is larger than $1$.
        \item When $CN-10 < 0$, the derivative of $\Bar{h}_{CN}(y)$ is
        \begin{align*}
            \Bar{h}'_{CN}(y) = 3(CN-10)y^2 - 2(CN-13)y -6.
        \end{align*}
        Let $\Bar{h}'_{CN}(y) = 0$, we obtain two roots, one is $x_1 = \frac{2(CN-13) + \sqrt{4(CN-13)^2 + 72(CN-10)}}{6(CN-10)}$, and the other one is $x_2 = \frac{2(CN-13) - \sqrt{4(CN-13)^2 + 72(CN-10)}}{6(CN-10)}$. Look at the term in the square root, $4(CN-13)^2 + 72(CN-10)= 4C^2N^2 - 32CN -44$. It has one negative $4-3\sqrt{3} \approx -1.1962$ and one positive root $4 + 3 \sqrt{3}\approx 9.1692$. Thus, $\Bar{h}'_{CN}(y)$ does not have real roots when $1 < CN < 10$, and hence $\Bar{h}_{CN}(y)$ is always a decreasing function in this case. Since $\Bar{h}_{CN}(0)=1>0$, $\Bar{h}_{CN}(1)=-2<0$, $\Bar{h}_{CN}(y)$ has only one root in $[0,1]$.
    \end{itemize}

    Now, we prove that $\Bar{h}'_{CN}(y) > 0$ for $y=0$ and $y=\frac{1}{N}$ when $N>C+4$. For $y=0$, $\Bar{h}_{CN}(0)=1>0$. When $y=\frac{1}{N}$,
    \begin{align*}
        \Bar{h}_{CN}(\frac{1}{N}) &= \frac{CN-10}{N^3} - \frac{CN-13}{N^2} - \frac{6}{N} + 1 \\
        & = \frac{N^3-(6+C)N^2+(13+C)N-10}{N^3}.
    \end{align*}
    When $N > C+4$,
    \begin{align*}
        \Bar{h}_{CN}(\frac{1}{N}) \geq& \frac{(C+5)^3-(6+C)(C+5)^2+(13+C)(C+5)}{N^3} \\
        &  - \frac{10}{N^3} =  \frac{8C+30}{N^3} > 0.
    \end{align*}

    Therefore, the two claims for $\beta$ and $\Bar{h}_{CN}(y)$ have been proved. So, $\beta>\frac{1}{N}$, when $N>C+4$.

%{\appendix[Proof of the Zonklar Equations]
%Use $\backslash${\tt{appendix}} if you have a single appendix:
%Do not use $\backslash${\tt{section}} anymore after $\backslash${\tt{appendix}}, %only $\backslash${\tt{section*}}.
%If you have multiple appendixes use $\backslash${\tt{appendices}} then use %$\backslash${\tt{section}} to start each appendix.
%You must declare a $\backslash${\tt{section}} before using any %$\backslash${\tt{subsection}} or using $\backslash${\tt{label}} %($\backslash${\tt{appendices}} by itself
% starts a section numbered zero.)}


\newpage

\bibliographystyle{ieeetr}
\bibliography{Main}
 
\vspace{11pt}

\begin{IEEEbiography}[{\includegraphics[width=1in,height=1.25in,clip,keepaspectratio]{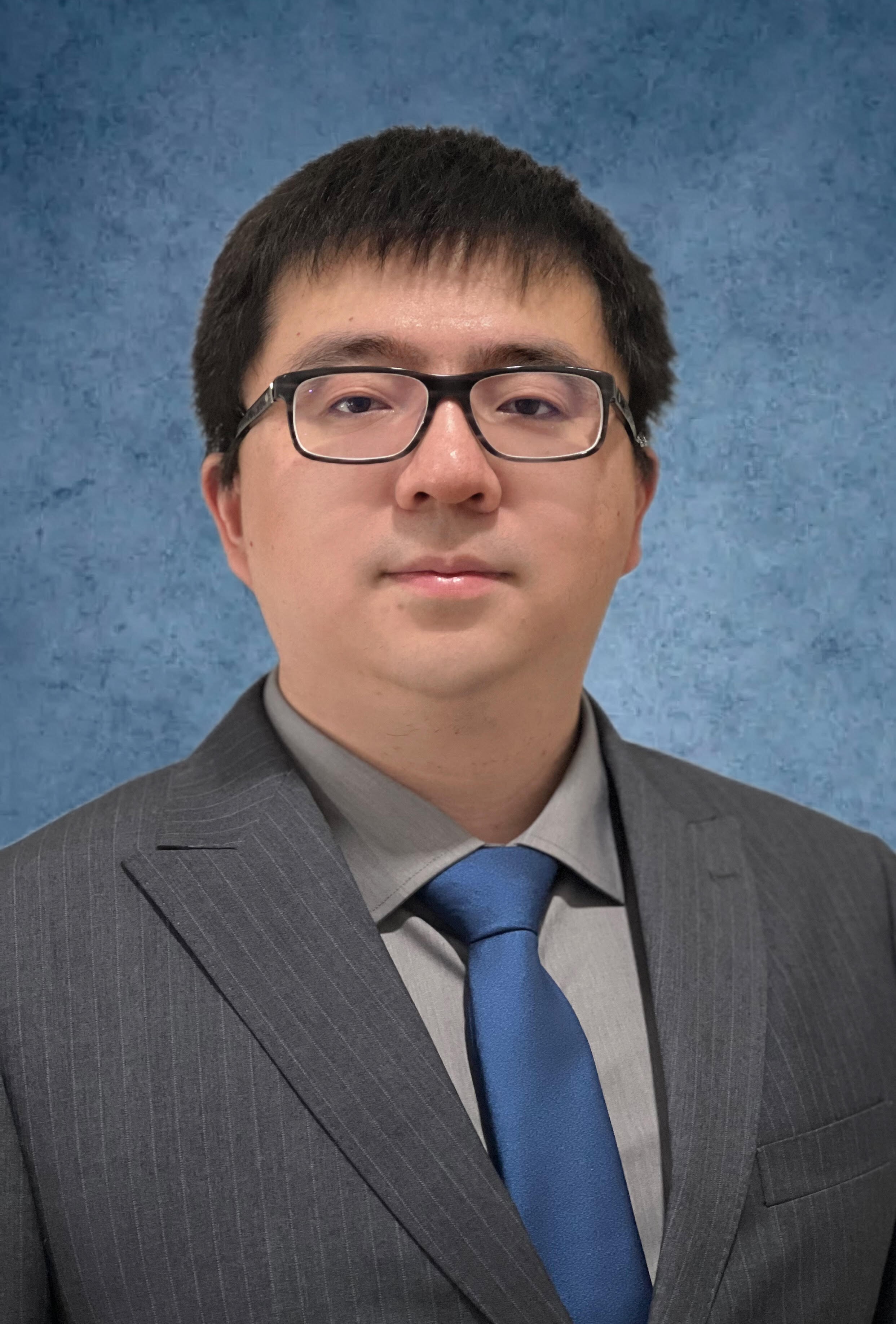}}]{Siqi Fan}
(Graduate Student Member, IEEE) received his B.S. degree in Microelectronics from Shanghai Jiao Tong University, Shanghai, China, in 2016, and his M.S. degree in Electrical Engineering from Texas A\&M University, Texas, United States, in 2019. Since 2020, he has been pursuing a Ph.D. degree in Computer Engineering at Texas A\&M University. His research interests include wireless communication, edge computing, and stochastic system optimization. He has recent publications in IEEE INFOCOM, ISIT, and ICC.
\end{IEEEbiography}

\vspace{11pt}

\begin{IEEEbiography}[{\includegraphics[width=1in,height=1.25in,clip,keepaspectratio]{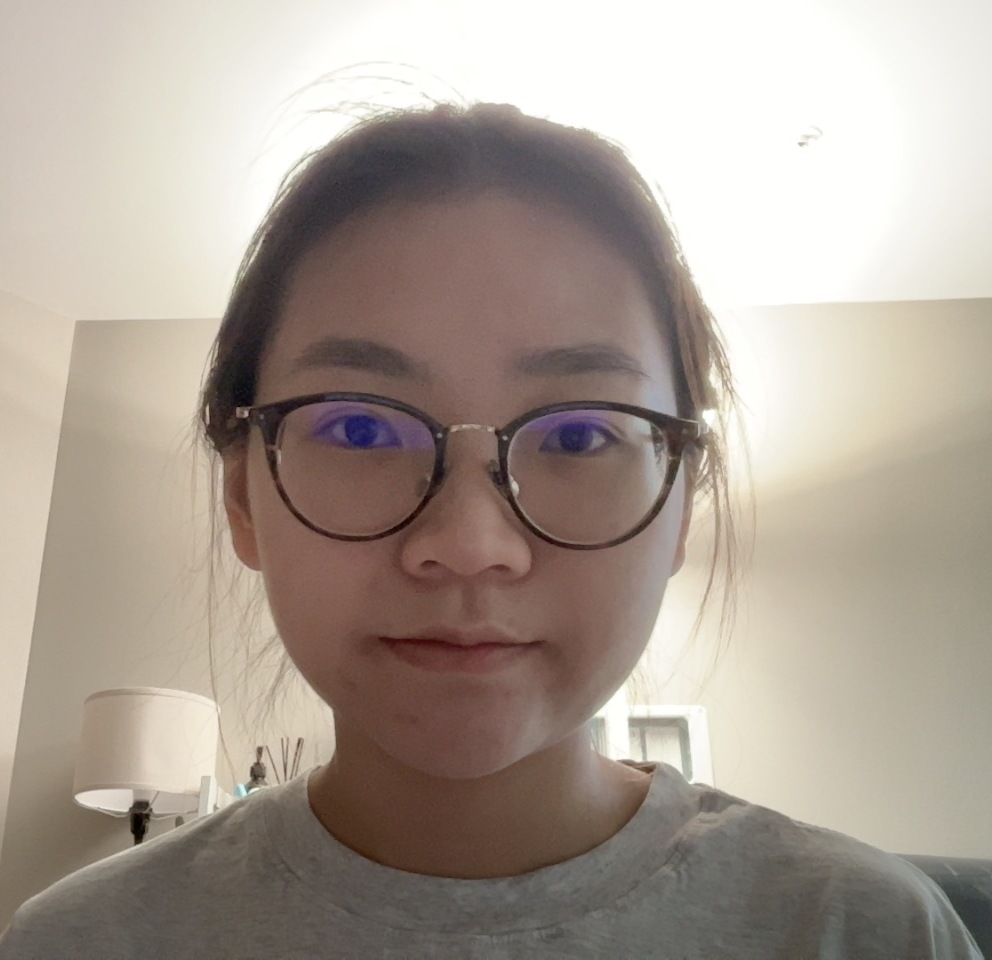}}]{Yuxin Zhong} is an Electrical and Computer Engineering Ph.D. student at Georgia Institute of Technology. She received a bachelor's degree in Electrical Engineering from Jilin University, China and a master's degree in Electrical and Computer Engineering from the University of Michigan, Ann Arbor. She is interested in privacy control in distributed sensor infrastructure, energy optimization in distributed sensor networks, and system optimization.
\end{IEEEbiography}

\vspace{11pt}

\begin{IEEEbiography}[{\includegraphics[width=1in,height=1.25in,clip,keepaspectratio]{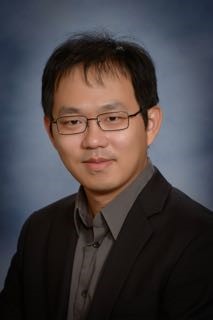}}]{I-Hong Hou}
(Senior Member, IEEE) is an Associate Professor in the ECE Department of the Texas A\&M University. He received his Ph.D. from the Computer Science Department of the University of Illinois at Urbana-Champaign. His research interests include wireless networks, edge/cloud computing, and reinforcement learning. His work has received the Best Paper Award from ACM MobiHoc 2017 and ACM MobiHoc 2020, and Best Student Paper Award from WiOpt 2017. He has also received the C.W. Gear Outstanding Graduate Student Award from the University of Illinois at Urbana-Champaign, and the Silver Prize in the Asian Pacific Mathematics Olympiad.
\end{IEEEbiography}

\vspace{11pt}

\begin{IEEEbiography}[{\includegraphics[width=1in,height=1.25in,clip,keepaspectratio]{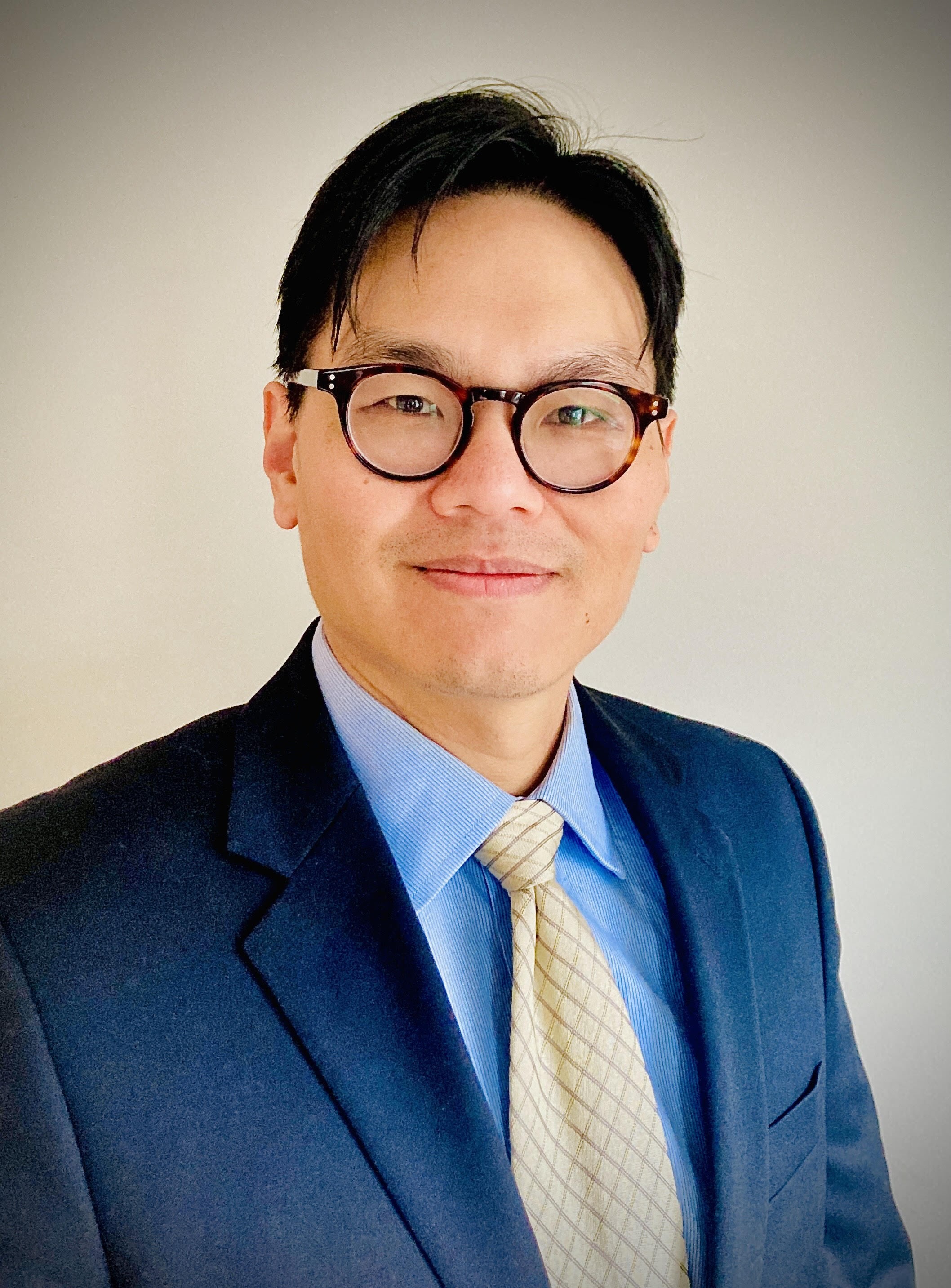}}]{Clement Kam}
(Senior Member, IEEE) received the BS degree in electrical and computer engineering from Cornell University, Ithaca, NY, in 2001, and the M.S. and Ph.D. degrees in electrical engineering from the University of California, San Diego, in 2006 and 2010, respectively. He is currently the Head of the Wireless Network Theory Section at the U.S. Naval Research Laboratory, Washington, DC. His research interests include ad hoc networks, cross-layer design, Age of Information, Semantics of Information, and reinforcement learning.
\end{IEEEbiography}



\vfill

\end{document}